\documentclass[reprint,twocolumn,aps,prb,showpacs,longbibliography]{revtex4-1}

\usepackage{amsmath}
\usepackage{amssymb}
\usepackage{graphicx}
\usepackage{dcolumn}
\usepackage{bm}
\usepackage[colorlinks=true, allcolors=blue]{hyperref}
\usepackage{epstopdf}
\usepackage[utf8]{inputenc}
\usepackage{amssymb}
\begin{document}

\title{Spontaneously broken chiral symmetry in the interacting Kane-Mele model}

\author{Minghuan Zeng$^{1}$}
\author{Junjie Zeng$^{1}$}
\author{Ling Qin$^{2}$}
\author{Shiping Feng$^{3,4}$}
\author{Donghui Xu$^{1,5}$}
\email{donghuixu@cqu.edu.cn}
\author{Rui Wang$^{1,5}$}
\email{rcwang@cqu.edu.cn}

\affiliation{$^{1}$Institute for Structure and Function \& Department of Physics \& Chongqing Key Laboratory for Strongly Coupled Physics, Chongqing University, Chongqing, 400044, P. R. China}

\affiliation{$^{2}$College of Physics and Engineering, Chengdu Normal University, Chengdu, 611130, Sichuan, China}

\affiliation{$^{3}$Department of Physics, Faculty of Arts and Science, Beijing Normal University, Zhuhai, 519087, China}

\affiliation{$^{4}$School of Physics and Astronomy, Beijing Normal University, Beijing, 100875, China}

\affiliation{$^{5}$Center of Quantum materials and devices, Chongqing University, Chongqing 400044, P. R. China}

\begin{abstract}
The essential properties of the half-filled interacting Kane-Mele model on a hexagon
lattice is studied using the slave rotor approach. It is shown clearly that a long-range
charge-order state with spontaneously broken chiral symmetry emerges in the weak and
moderate interaction regimes, as well as a presumed site-selected topological Mott
insulator state in the stronger interaction regime with $U<U_{\rm Mott}$, where
$U_{\rm Mott}$ is the critical interaction strength, and in the case of $U>U_{\rm Mott}$,
the system is transited into the usual topological Mott state. This new charge-order
state has lower energy compared to the usual topological band insulator (TBI) state with
chiral symmetry, and thus is named as non-chiral TBI state. More specifically, in this
non-chiral TBI state without any long-range magnetic order, a long-range charge order with
different electron occupation on two sublattices appears in the absence of
external sublattice field. The spontaneously broken chiral symmetry gives rise to a
special helical edge state, which has different spin accumulation on opposite edges of the
cylinder with periodic boundary condition in the zigzag direction, and thus leads to a net
spin current across the system. This net spin current would be further strengthened if the
nearest neighbor electron Coulomb interaction is taken into account as well, because it is
favorable for the long-range charge order with different electron occupation on sublattices.
\end{abstract}


\maketitle


\section{Introduction}

In the early days of the quantum Hall effect research, Haldane proposed that the integer
quantum Hall effect without Landau levels may exist in a time reversal symmetry broken
system with zero net magnetic flux through the unit cell in a two-dimensional(2D) graphene
system\cite{Haldane88}. Later, a notable model was introduced by Kane and
Mele\cite{Kane0509,Kane0511} to capture the essential physics of the quantum spin Hall
(QSH) effect, which
consists of two copies with opposite sign for up and down spins of Haldane model that is
connected by the time reversal symmetry. In this Kane-Mele model\cite{Kane0509,Kane0511},
a new type of Z$_{2}$
topological invariant is introduced to characterize the topologically nontrivial state.
Most importantly, although the charge accumulation on both sides of the device
vanishes because electrons with up and down spin move in the opposite direction on the
same side, the opposite spin accumulation is realized in the QSH state, which thus
manifests its importance in designing the spintronics devices. Since the proposal
that the QSH effect may be realized in the Graphene material\cite{Kane0509,Kane0511},
some researchers strive to look for QSH materials,
such as graphene\cite{Novoselov05,Zhang05,Yao07}, silicene\cite{Liu1111,Chowdhury16}, Germanium\cite{Liu1108},
black arsenic\cite{Sheng21,Wang18}, etc., however, only few of these
materials has been verified experimentally for the existence of the QSH effect. Recently,
the QSH effect has been observed experimentally in the AB-moire-stacked transition
metal dichalcogenide (TMD) bilayer MoTe$_{2}$/WSe$_{2}$\cite{Li21,zhao22,Tschirhart23,Zui24},
and has been predicted by the first-principles band calculation in a transitional metal oxide Na$_{2}$IrO$_{3}$~\cite{Shitade09} and monolayer
TMDs MX$_{2}$ in 1T' structure with M=(Mo, W) and X=(S, Se, Te)\cite{Qian14}, where both
the spin-orbit interaction and electron correlation play the key roles. On the other
hand, many theoretical
works\cite{Feldner10,Hutchinson21,Rademaker22,Devakul22,Ghorbani23,Qiu23,Guerci24,Wagner24,Gupta2024}
deal with the spin-orbit coupling and electron Coulombic interaction
on the same footing, and some results about the spin density wave and topological states
have been obtained, where the spontaneous edge magnetism with non-magnetic bulk on a nanoribbon
periodic in the zigzag direction is predicted based on mean-field approximations\cite{Ghorbani23,Guerci24},
which is also consistent with experimental observations on narrow graphene nanoribbons~\cite{Magda14}. However, the majority of these previous studies focus on the time reversal symmetry broken states because of the spontaneous magnetization at large interaction strengths~\cite{Devakul22,Guerci24} which drives the system from the topological insulator to the Chern insulator,
while the electron interaction can not only lead to the spin density wave, but also the charge density wave, while how the latter affects the electron topological properties has not been discussed so far.

Here we start from the interacting Kane-Mele model on a hexagon lattice with two atoms per unit cell and adopt the slave-rotor method~\cite{Florens02,Florens04,Rachel10,Pesin10,Wagner24} to study the onsite coulomb interaction effects on topological properties such as the electron dispersion for a bulk system with periodic boundary conditions and a nanoribbon periodic in the zigzag direction, where we focus on the new topological band insulator(TBI) state with a long-range charge order in the system. The slave-rotor representation is proposed to solve multiorbital Anderson quantum impurity model~\cite{Florens02} first, and then is extended to study the metal-insulator Mott transition for strongly correlated systems~\cite{Florens04}. The slave-rotor representation captures the correct physics in the metallic and insulating state at small and large interaction strengths, respectively, as well as the transition between them, namely a Brinkman-Rice transition at commensurate fillings with a vanishing renormalization factor~\cite{Brinkman70},
and opening a charge gap in the electron density of states corresponding to the Mott-Hubbard physics~\cite{Hubbard63,Hubbard64}.
In addition, the slave-rotor representation is more economical computationally than the conventional slave-boson representation~\cite{Kotliar86} which requires 4 slave bosons for each flavor, while for the former
researchers only need to introduce one constrained bosonic field $X$ with $|X|^2=1$.

After using the slave rotor method to systematically study the interacting Kane-Mele model, we find a new type of TBI state with spontaneously broken chiral symmetry which has lower energy compared to
the conventional TBI state maintaining the chiral symmetry in the presence of intrinsic spin-orbit coupling(SOC) and on-site coulomb interaction between electrons, which is thus named by non-chiral TBI state.
In addition, there exists a long range charge order with different electron
occupation on two sublattices in this non-chiral TBI state, which not only leads to unequal electron quasiparticle weight located on sublattice A and B as a result of different renormalization factors on them
for the bulk system, but also a special helical edge state with inequivalent spin accumulation located
at opposite edges of the cylinder periodic in the zigzag direction, and then gives rise to a net spin current across this system.
Most importantly, the net spin accumulation across the above cylinder because of the spontaneously broken chiral symmetry
can be further strengthened by the nearest neighbor coulomb interaction between electrons, and the related results will be presented else where.

 The rest of this paper is organized as follows. In Sec. \ref{Formalism}, we introduce the interacting Kane-Mele model which consists of the electron hopping between nearest neighbor sites, the intrinsic SOC,
 and the on-site coulomb interaction between electrons, as well as the slave rotor representation.
 The results are shown in Sec. \ref{Result}, where we find a novel TBI state with the spontaneously broken chiral symmetry in the presence of intrinsic SOC and on-site coulomb interaction between electrons.
 Moreover, the physical properties
 of the bulk system with periodic boundary conditions and the nanoribbon periodic in the zigzag direction are thoroughly investigated. Finally, we give a summary in Sec. \ref{conclude}. In addition, the details for the lattice setup in real and momentum space, the theoretical derivation of the Green's functions and self-consistent equations, the free energy of the system, the Lagrange of the nanoribbon periodic in the zigzag direction and its Green's functions are given in Appendix \ref{Lattice setup}, \ref{GF-SC Eqs}, \ref{Free-En}, and \ref{Zigzag-Nanoribbon}, respectively.

\section{Theoretical Framework}\label{Formalism}

The Hamiltonian of the interacting Kane-Mele model, a.k.a. Kane-Mele-Hubbard model, consists of three parts and reads ${\rm H} = {\rm H}_{t}+{\rm H}_{\lambda}+{\rm H}_{U}$. Please note that we do not break the chiral symmetry by hand
via introducing the staggered potential to distinguish two sublattices, where ${\rm H}_{t}$, ${\rm H}_{\lambda}$, and ${\rm H}_{U}$ represent the electron hoping between the nearest neighbor sites
together with the chemical potential introduced to determine the electron filling, intrinsic spin-orbit coupling, and on-site electron coulomb interaction, respectively, which are explicitly expressed as
\begin{subequations}
\begin{eqnarray}
{\rm H}_{t}&=&-t\sum_{\langle ij\rangle}[C_{iA}^{\dagger}C_{jB}+\mathrm{H.C.} + \mu\delta_{ij} \sum_{s}C_{is}^{\dagger}C_{js}]\;, \\
{\rm H}_{\lambda}&=&\lambda\sum_{s=A,B}\sum_{\langle\langle ij\rangle\rangle}C_{is}^{\dagger}e^{i\tfrac{\pi}{2}\nu_{ij}\sigma^{z}}C_{js}\;, \\
{\rm H}_{U}&=&\frac{U}{2}\sum_{s=A,B}\sum_{i}(\sum_{\sigma}n_{is\sigma}-1)^2 \;,
\end{eqnarray}
\end{subequations}
where $\langle ij\rangle$ and $\langle\langle ij\rangle\rangle$ denote that the summation is over all the nearest and next nearest neighbor sites, respectively;
$C_{is}^{\dagger}=(C_{is\uparrow}^{\dagger},C_{is\downarrow}^{\dagger})$ is a two-component spinor with $s=A,B$ denoting two sublattices; $\sigma^{z}$ is the $z$ component of pauli matrices;
$\nu_{ij}=\pm 1$ is the Haldane factor for clockwise and anticlockwise path connecting the next nearest neighbor sites $i$ and $j$;
$n_{is\sigma}=C_{is\sigma}^{\dagger}C_{is\sigma}$ is the electron occupation number operator at site $is$ with spin $\sigma$. The nearest neighbor hoping integral $t$ and the lattice constant $a$ have been set
as the energy and length unit, respectively. Moreover, when compared to experiments, the nearest neighbor hopping magnitude is set as $t=1000$K.
The lattice setup in real space and its Brillouin zone in momentum space have been given in Appendix \ref{Lattice setup}.
Because we only deal with the half-filled system where the particle-hole symmetry requires that the chemical potential vanishes~\cite{Rachel10,Wagner24}, the chemical potential term in ${\rm H}_{t}$ is omitted in the following calculations.

In the slave rotor representation, an electron operator is decomposed as a direct product of a rotor specified by a phase degree of freedom $\theta_{\bm{r}s}$ which is conjugate
to the local charge represented by an angular momentum operator $L_{\bm{r}s}=-i\partial_{\theta_{\bm{r}s}}$ and an auxiliary fermion operator $f_{\bm{r}s\sigma}$ representing the spin degree of freedom(spinon)
\begin{equation}
C_{\bm{r}s\sigma}=e^{-i\theta_{\bm{r}s}}f_{\bm{r}s\sigma} \;,
\end{equation}
which is subjected to the local constraint $L_{\bm{r}s}-\sum_{\sigma}f_{\bm{r}s\sigma}^{\dagger}f_{\bm{r}s\sigma}+1=0$. Then the terms in Hamiltonian ${\rm H}$ can be
reexpressed as
\begin{subequations}\label{Ham-srotor}
\begin{eqnarray}
{\rm H}_{t}&=&-t\sum_{\langle ij\rangle}[f_{iA}^{\dagger}f_{jB}e^{i(\theta_{iA}-\theta_{jB})}+\mathrm{H.C.}]\;, \\
{\rm H}_{\lambda}&=&\lambda\sum_{s=A,B}\sum_{\langle\langle ij\rangle\rangle}f_{is}^{\dagger}e^{i(\theta_{is}-\theta_{js})}
e^{i\tfrac{\pi}{2}\nu_{ij}\sigma^{z}}f_{js}\;, \\
{\rm H}_{U}&=&\frac{U}{2}\sum_{s=A,B}\sum_{i}{L_{is}}^2 \;,
\end{eqnarray}
\end{subequations}
where the two-component spinor $f_{is}^{\dagger}$ representing the spin degree of freedom reads $(f_{is\uparrow}^{\dagger},f_{is\downarrow}^{\dagger})$. We deal with this interacting Kane-Mele model in the saddle
point approximation which has been justified in the large spin degeneracy limit~\cite{Cox93,Florens02,Florens04}, while is still reliable to obtain reasonable results for low spin degeneracy systems~\cite{Pesin10,Rachel10,Wagner24}. The emergent U(1) gauge field~\cite{Lee92,Lee05,Senthil08} as a consequence of the slave particle representation towards strongly correlated
systems is not taken into account. However T. Senthil has treated the U(1) gauge field due to the redundancy in the slave rotor representation of electron operators seriously~\cite{Senthil08},
and found that the gauge interaction only leads to the analytic correction to the boson self-energy and thus does not change the critical singularities coming from the boson interaction.
In addition, Dmytro Pesin and Leon Balents argued that the U(1) gauge fluctuations beyond the mean-field fixed point only lead to the continuous transition changing into the first order class~\cite{Pesin10}.
Technical details of the derivation for the spinon and chargon Green's function, $G_{f\sigma}^{ss'}(i-j,\tau)=-\langle T_{\tau}f_{is\sigma}(\tau) f_{js'\sigma}^{\dagger}(0)\rangle$ and $G_{X}^{s's}(j-i,0-\tau)=
\langle T_{\tau}X_{js'}(0)X_{is}^{\ast}(\tau)\rangle$, respectively, as well as the self-consistent equations used to determine the mean-field parameters appearing when we deal with the Hamiltonian \eqref{Ham-srotor}
in the saddle point approximation are given in Appendix \ref{GF-SC Eqs}, where $X_{is}(\tau)=e^{i\theta_{is}(\tau)}$ such that $|X_{is}|^2=1$.

\section{Results}\label{Result}

The coulomb interaction between electrons plays an important role in determining the rich physics of strongly correlated systems, especially some novel states with different spontaneously broken symmetry such as antiferromagnets appearing
in the deeply underdoped regime, the superconducting state appearing around the optimal doping, and et al~\cite{Zaanen15}. Though the interacting KM model has been studied for almost twenty years, while whether the electron coulomb interaction together with the quantum geometry introduced by intrinsic SOC could induce some new states with spontaneously broken symmetry is unclear.
In this work, we adopted the slave rotor method to systematically investigate the half-filled interacting KM model for small and moderate interaction strengths, and find a novel topological band insulator state with spontaneously broken chiral symmetry which leads to different electron occupation
on two sublattices in the absence of any long-range magnetic order. We further study the physical properties of the edge state on a nanoribbon periodic in the zigzag direction, and find that compared to the conventional TBI state
where the spin current localized at opposite boundaries offsets each other, a net spin current across this nanoribbon survives as a natural consequence of the spontaneously broken chiral symmetry, which provides further
guidance for designing more advanced spintronics devices.

\subsection{The self-consistent parameters and phase diagram in $\lambda$-$U$ parameter space }

From the poles of the bosonic chargon Green's function $G_{X}^{s's}(\bm{k},i\nu)$[See Eq.\eqref{GF-Psi-X-2} in Appendix \ref{GF-SC Eqs}] with $i\nu=i2n\pi/\beta$ being the bosonic Matsubara frequency where $n$ is integers,
and $\beta$ is the inverse of temperature, i.e., $\beta=1/T$, we can obtain the energy dispersion $\omega(\bm{k})$ for the constrained bosonic field $X$:
\begin{eqnarray}
&&\omega(\bm{k}) = \pm\sqrt{\frac{U}{2}}\Big\{\rho_{A}+Q_{AA}^{X}\epsilon_{\bm{k}}+\rho_{B}+Q_{BB}^{X}\epsilon_{\bm{k}}\pm \nonumber\\
&& \sqrt{[\rho_{A}+Q_{AA}^{X}\epsilon_{\bm{k}}-\rho_{B}-Q_{BB}^{X}\epsilon_{\bm{k}}]^2+4|Q_{AB}^{X}V_{\bm{k}}|^2 }   \Big\}^{1/2}\;,
\end{eqnarray}
where $\rho_{A/B}$ is the Lagrange multiplier to implement the constraint $|X_{A/B}|^2=1$, respectively; $V_{\bm{k}}=\sum_{\bm{u}}e^{i\bm{k}\cdot\bm{u}}$ as well as $\epsilon_{\bm{k}}=2\lambda\sum_{j=1,2,3}\cos(\bm{k}\cdot\bm{\gamma}_{j})$
are given in Appendix \ref{Lattice setup} with the nearest neighbor vector $\bm{u}$ and next nearest neighbor vector $\bm{\gamma}_{j}$; And
\begin{subequations}
\begin{eqnarray}
Q_{ss}^{X}&=&\frac{1}{6N\beta}\sum_{\langle\langle ij\rangle\rangle}\int_{0}^{\beta}d\tau \langle f_{is}^{\dagger}e^{i\tfrac{\pi}{2}\nu_{ij}\sigma^{z}}f_{js}\rangle \;,\\
Q_{AB}^{X}&=&\frac{1}{3N\beta}\sum_{\langle ij\rangle}\int_{0}^{\beta}d\tau \langle f_{iA}^{\dagger}f_{jB}\rangle\;,
\end{eqnarray}
\end{subequations}
where $N$ is the lattice size. In the slave-rotor representation, the bosonic field $X$ is condensed and exhibits a long-range order in the metallic state along with nonvanishing quasiparticle weight
$Z_{ss'}=\langle X_{s}(\bm{k}=\bm{0},\tau)X_{s'}^{\dagger}(\bm{k}=\bm{0},\tau)\rangle/N$, which then requires the gapless bosonic energy dispersion $\omega(\bm{k})$ at $\bm{k}=\bm{0}$.
Thus there exist two solutions in the TBI state with different Lagrange multiplier $\rho_{s}$:
\begin{subequations}
\begin{eqnarray}
\rho_{A}&=&\rho_{B}=Q_{AB}^{X}D-6\lambda Q_{AA}^{X}\;,\label{sym-sol} \\
\rho_{B} &=& \frac{(Q_{AB}^{X}D)^2}{\rho_{A}+6\lambda Q_{AA}^{X}}-6\lambda Q_{BB}^{X} \label{nonsym-sol}
\end{eqnarray}
\end{subequations}
with $D$ being the half bandwidth of the system with zero SOC strength. The solution in Eq.\eqref{sym-sol} corresponds to the conventional topological band insulator state with chiral symmetry which has been investigated
before~\cite{Rachel10,Wagner24}, and this state is renamed by the chiral TBI state in the following. However, the other solution with $\rho_{A}\neq \rho_{B}$ reveals a novel TBI state with spontaneously broken chiral symmetry
in the presence of on-site electron coulomb interaction and intrinsic SOC which is clearly demonstrated by the electron energy dispersion of the bulk system with periodic boundary conditions and the cylindrical nanoribbon periodic in the zigzag
direction as shown in Sec. \ref{bulk-dispersion} and \ref{Edge-Modes-EL}, respectively, and this new state is thus referred to as non-chiral TBI state.
Moreover, to return to the solution of Hubbard model with vanishing SOC strength, the Lagrange multiplier on sublattice A is set as $\rho_{A}=Q_{AB}^{X}D$. We have also tried other forms of $\rho_{A}$
as shown in Appenx \ref{GF-SC Eqs}, and found that different $\rho_{A}$'s only lead to some quantitative changes to the self-consistent parameters,
then the physics of this novel TBI state with $\rho_{A}\neq \rho_{B}$.

\begin{figure}[h!]
\centering
\includegraphics[scale=0.75]{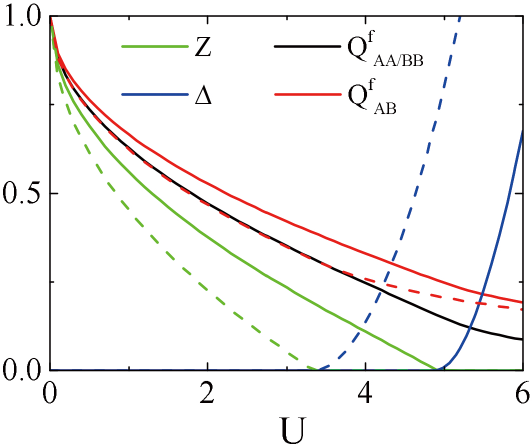}
\caption{(Color online) The quasiparticle renormalization factor $Z$, charge gap $\Delta$, and the mean-field parameters $Q_{ss'}^{f}$ in the spin section with $s(s')$ being the sublattice index A and B, respectively,
at $\lambda=0.5$ and $T=1{\rm K}$ as a function of interaction strength $U$ for the TBI state with chiral symmetry. The above quantities at $\lambda=0$ are shown by dashed lines in the corresponding color.\label{Paras-Chiral-Sym}}
\end{figure}
Before exploring the physics in the new TBI state with $\rho_{A}\neq \rho_{B}$, we first restudy the quasiparticle renormalization factor $Z$, charge gap $\Delta$,
and the mean-field parameters $Q_{ss'}^{f}$[c.f. Appendix \ref{GF-SC Eqs}] in the TBI state with chiral symmetry at $\lambda=0.5$
and $T=1{\rm K}$ as a function of interaction strength $U$, where the charge gap is obtained according to the bosonic dispersion $\omega_{3}(\bm{k})$,
which can be extracted by setting $\rho_{A}=\rho_{B}=\rho$ and $Q_{AA}^{X}=Q_{BB}^{X}=Q_{ss}^{X}$
\begin{eqnarray}
\omega_{3}(\bm{k})&=&\sqrt{U[\rho+Q_{ss}^{X}\epsilon_{\bm{k}}-Q_{AB}^{X}|V_{\bm{k}}|]} \nonumber\\
&=& \sqrt{\Delta+U[Q_{ss}^{X}(\epsilon_{\bm{k}}-\epsilon_{\bm{0}})-Q_{AB}^{X}(|V_{\bm{k}}|-|V_{\bm{0}}|)]}
\end{eqnarray}
with $\Delta=U[\rho+Q_{ss}^{X}\epsilon_{\bm{0}}-Q_{AB}^{X}|V_{\bm{0}}|]$.
From Fig.\ref{Paras-Chiral-Sym}, the intrinsic SOC does not change the physics of the metal-insulator Mott transition substantially
where the quasiparticle renormalization factor vanishes and a charge gap is opened, except that the critical interaction strength $U_{\rm Mott}$ responsible for this metal-insulator Mott transition is increased
monotonically by the growing SOC strength compared to $U_{\rm Mott}\approx 3.3t$ at zero SOC strength which is denoted by the intersection of the green and blue dashed lines shown in this figure.
Please note that our results shown in Fig.\ref{Paras-Chiral-Sym} are somewhat different from Fig.S2 in Niklas Wagner, et al.'s work~\cite{Wagner24}, while we hold that
these subtle differences should be caused by different temperatures which in their work is set as $T=10{\rm K}$, ten times larger than ours,
and different lattice size which in our work is set as $N=400 \times 400$.

\begin{figure}[h!]
\centering
\includegraphics[scale=0.6]{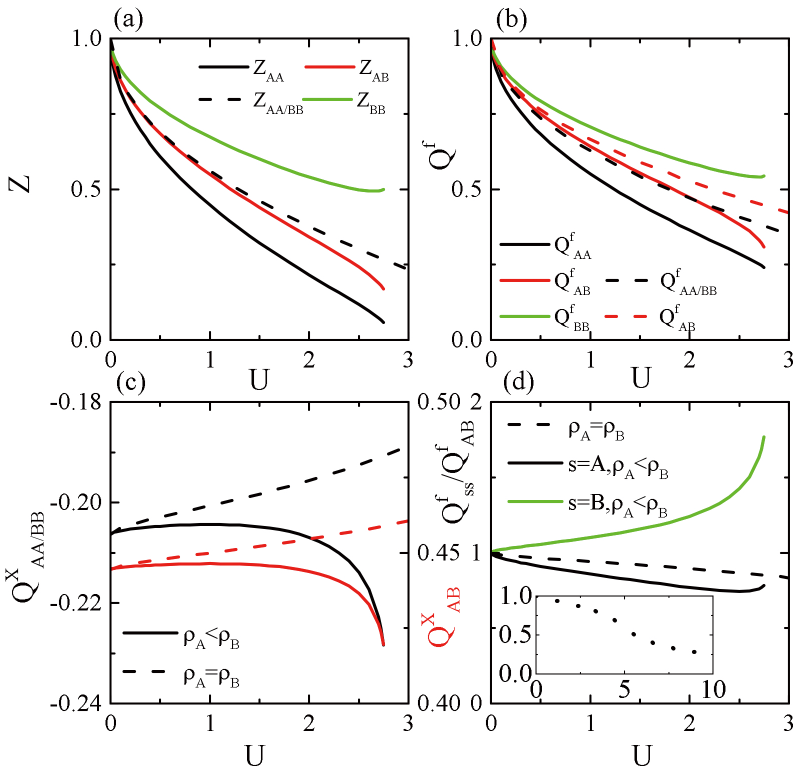}
\caption{(Color online) (a)The quasiparticle renormalization factor $Z_{ss'}$, the mean-field parameters (b)$Q_{ss'}^{f}$ in the spin section and (c)$Q_{ss'}^{X}$ in the charge section,
as well as (d)the effective SOC strength $Q_{ss}^{f}/Q_{AB}^{f}$ as a function of interaction strength $U$ at $\lambda=0.5$
and $T=1{\rm K}$ in the TBI state with(dashed lines) and without(solid lines) chiral symmetry. Inset in Fig.\ref{Paras-Chiral-Sym-NSym}d shows the effective intrinsic SOC strength
as a function of onsite coulomb interaction strength in the chiral TBI state with $0<U<10$. \label{Paras-Chiral-Sym-NSym}}
\end{figure}
We are now ready to study the dependence of the self-consistent parameters $Z$, $Q_{ss'}^{f}$, and $Q_{ss'}^{X}$ upon the interaction strength $U$ in the presence of intrinsic SOC, and these quantities as a function of
interaction strength $U$ at $\lambda=0.5$ are plotted in Fig.\ref{Paras-Chiral-Sym-NSym}, where the dashed and solid lines represent the results in the chiral and non-chiral TBI state, respectively.
In Fig.\ref{Paras-Chiral-Sym-NSym}(a), $Z$ in the chiral TBI state is located between the quasiparticle renormalization factors $Z_{AA}$ and $Z_{BB}$ in the non-chiral TBI state defined on sublattice
A and B, respectively, and they decrease monotonously as the interaction strength $U$ is raised. Though $Z_{ss}$ in the non-chiral TBI state exhibit similar behaviors as the chiral TBI state,
they deviate from the chiral TBI state in the presence of intrinsic SOC at infinitesimal interaction strengths, and their separation is enlarged as $U$ rises,
which then becomes pronounced at large interaction strengths. Thus the interacting KM model is presumed to transit into the site-selected topological Mott insulator(TMI) state
with $Z_{\rm A} = 0$ and $Z_{\rm B} > 0$
as the interaction strength further increases, though there is no direct evidence supporting this argument.
Moreover, the unequal $Z_{ss}$ with $s=$ A and B, respectively, lead to different electron quasiparticle weight localized on two sublattices, then the simplest charge order;
In Fig.\ref{Paras-Chiral-Sym-NSym}(b), $Q_{ss'}^{f}$ exhibit similar behaviors to the quasiparticle renormalization factor $Z_{ss'}$ while with higher values which is consistent with the results of Hubbard model\cite{Florens04}. Compared to $Z_{ss'}$ and $Q^{f}_{ss'}$, the parameter $Q^{X}_{ss'}$ in the chiral and non-chiral TBI state shown in Fig.\ref{Paras-Chiral-Sym-NSym}(c) behave in an opposite way, i.e.,
$Q^{X}_{ss'}$ in the chiral TBI state increases almost linearly as $U$ is raised, while in the non-chiral TBI state, $Q^{X}_{ss'}$ exhibits a nonlinear behavior, i.e., first increases moderately in the weak interaction limit,
then decreases monotonically as $U$ further increases, and drops dramatically in the vicinity of the phase boundary between the non-chiral TBI state and the site-selected TMI state. After implementing the charge-spin separation,
the Lagrange of interacting KM model is rewritten in the slave-rotor representation as Eq. \eqref{Lagrange-LargeM}, where the topological physics
of the non-interacting KM model is now inherited by the spinon component of Lagrange \eqref{Lagrange-LargeM} with the renormalization coefficients $Q_{ss}^{f}$ and $Q_{AB}^{f}$, then an effective SOC strength can be
defined as $\lambda_{{\rm eff}}=\lambda(Q_{ss}^{f}/Q_{AB}^{f})$, thus the renormalization effect on the SOC strength imposed by the electron coulomb interaction can be quantified by $Q_{ss}^{f}/Q_{AB}^{f}$, and its dependence
on interaction strength $U$ at $\lambda=0.5$ is presented in Fig.\ref{Paras-Chiral-Sym-NSym}(d). As shown by the black and red solid line, respectively, the renormalization factor $Q_{ss}^{f}/Q_{AB}^{f}$ on sublattice A and B behave in an opposite manner, where $Q_{AA}^{f}/Q_{AB}^{f}$ decreases monotonically as $U$ rises which is consistent with the chiral TBI state as shown by the black dashed line, while the other on sublattice B increases monotonically with an unexpected surge near the boundary between the non-chiral TBI state and site-selected TMI state. In addition, as shown by the black dashed line in the inset of this figure, the renormalization factor $Q_{ss}^{f}/Q_{AB}^{f}$ in the chiral TBI state decreases monotonically as $U$ increases, and in particular, this quantity drops significantly as the critical strength $U_{{\rm Mott}}$ responsible for the metal-insulator
Mott transition is reached.

\begin{figure}[h!]
\centering
\includegraphics[scale=0.5]{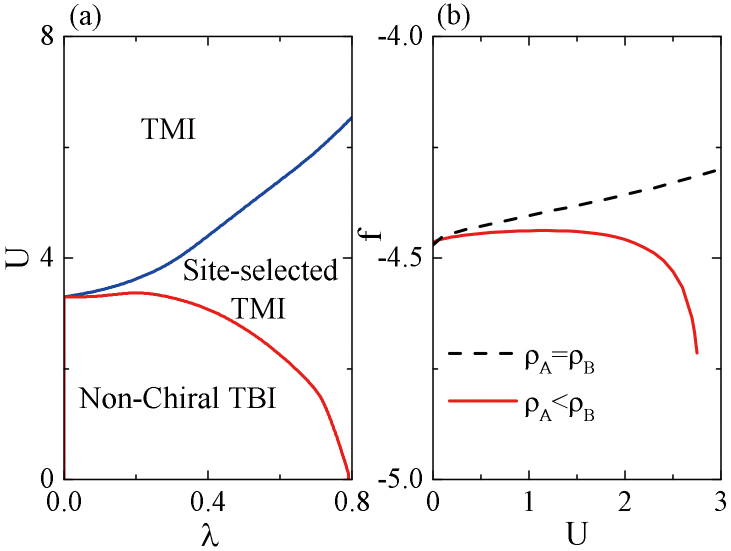}
\caption{(Color online) (a)The phase diagram of the interacting Kane-Mele model in the $\lambda-U$ parameter space which is divided into three regimes:
the topological band insulator with the spontaneously broken chiral symmetry(Non-chiral TBI), the site-selected topological Mott insulator (Site-selected TMI), and the topological Mott insulator (TMI) state at large interaction strengths;
(b)the free energy density $f=F/N$ as a function of the onsite interaction strength $U$ at $T=1$K and $\lambda=0.5$ where the black dashed line and red solid line represent
the free energy density in the chiral and non-chiral TBI state, respectively.\label{PD-FFig}}
\end{figure}
After systematic investigations, the phase diagram in the $\lambda-U$ parameter space is shown in Fig.\ref{PD-FFig}(a),
where the weak and moderate interaction regime is divided into two parts due to the spontaneously broken chiral symmetry: non-chiral TBI state and site-selected TMI state which are delineated by the red solid line.
The topological Mott insulator state has recently been thoroughly
studied by Niklas Wagner, et al.~\cite{Wagner24}, and they found that the topology of the electron Green's function zeros is identical to the poles of the spinon Green's function which had been proved theoretically~\cite{Blason23}. Our main finding is the existence of the new TBI state with spontaneously broken chiral symmetry at small and moderate interaction strengths,
and its phase boundary[marked by the red solid line] decreases monotonously with the
increment of the intrinsic SOC strength $\lambda$. Though there is no direct evidence supporting the existence of site-selected TMI state, as shown in Fig.\ref{Paras-Chiral-Sym-NSym}(a), the inequivalence between sublattice A and B in the non-chiral TBI state indicates that
the ground state of the interacting KM model as $U$ further increases ought to be the site-selected TMI state. Moreover, we have also adopted other approaches such as cluster slave spin method~\cite{Zeng21,Zeng22}
to further verify the existence
of site-selected TMI state, and the results will be presented else where. In Fig.\ref{PD-FFig}(b), the free energy density $f=F/N$[See Eq.\eqref{FE-Density} in Appendix \ref{Free-En}] in the chiral TBI state at $T=1$K and $\lambda=0.5$ increases linearly as the interaction strength $U$ rises before the Mott transition occurs, which is consistent with the Hubbard model where the SOC strength vanishes~\cite{Zeng21,Zeng22}. However in the non-chiral TBI state the free energy density
grows slowly first, while then decreases as the interaction strength is further enhanced, and drops dramatically as $U$ gets close to the phase boundary separating the non-chiral TBI state and site-selected TMI state. Most importantly,
the free energy density in the non-chiral TBI state is lower than the chiral TBI state which is more pronounced at large interaction strengths, indicating that the non-chiral TBI state is a potential candidate for the true ground state of the interacting KM model.

\subsection{The bulk energy dispersion in the chiral and non-chiral TBI state}\label{bulk-dispersion}

We in this subsection systematically investigate the bulk energy dispersion of the fermionic spinon field $f_{is\sigma}$, the constrained bosonic field $X_{is}$, and fermionic electron field $C_{is\sigma}$ in the chiral
and non-chiral TBI state, respectively, and their spectrum function are defined as
\begin{subequations}
\begin{eqnarray}
A_{f\sigma}^{ss'}(\bm{k},\omega) &=& -2{\rm Im} G_{f\sigma}^{ss'}(\bm{k},\omega+i0^{+}) \;, \\
A_{X}^{ss'}(\bm{q},\omega) &=& 2{\rm Im} G_{X}^{ss'}(\bm{q},\omega+i0^{+}) \;, \\
A_{\sigma}^{ss'}(\bm{k},\omega) &=& -2{\rm Im} G_{\sigma}^{ss'}(\bm{k},\omega+i0^{+})
\end{eqnarray}
\end{subequations}
with $G^{ss'}(\bm{k},\omega+i0^{+})$ being the retarded Green's function in momentum-energy space, and their counterparts in the Matsubara frequency representation have been given in Appendix \ref{GF-SC Eqs}.

\begin{figure}[h!]
\centering
\includegraphics[scale=0.3]{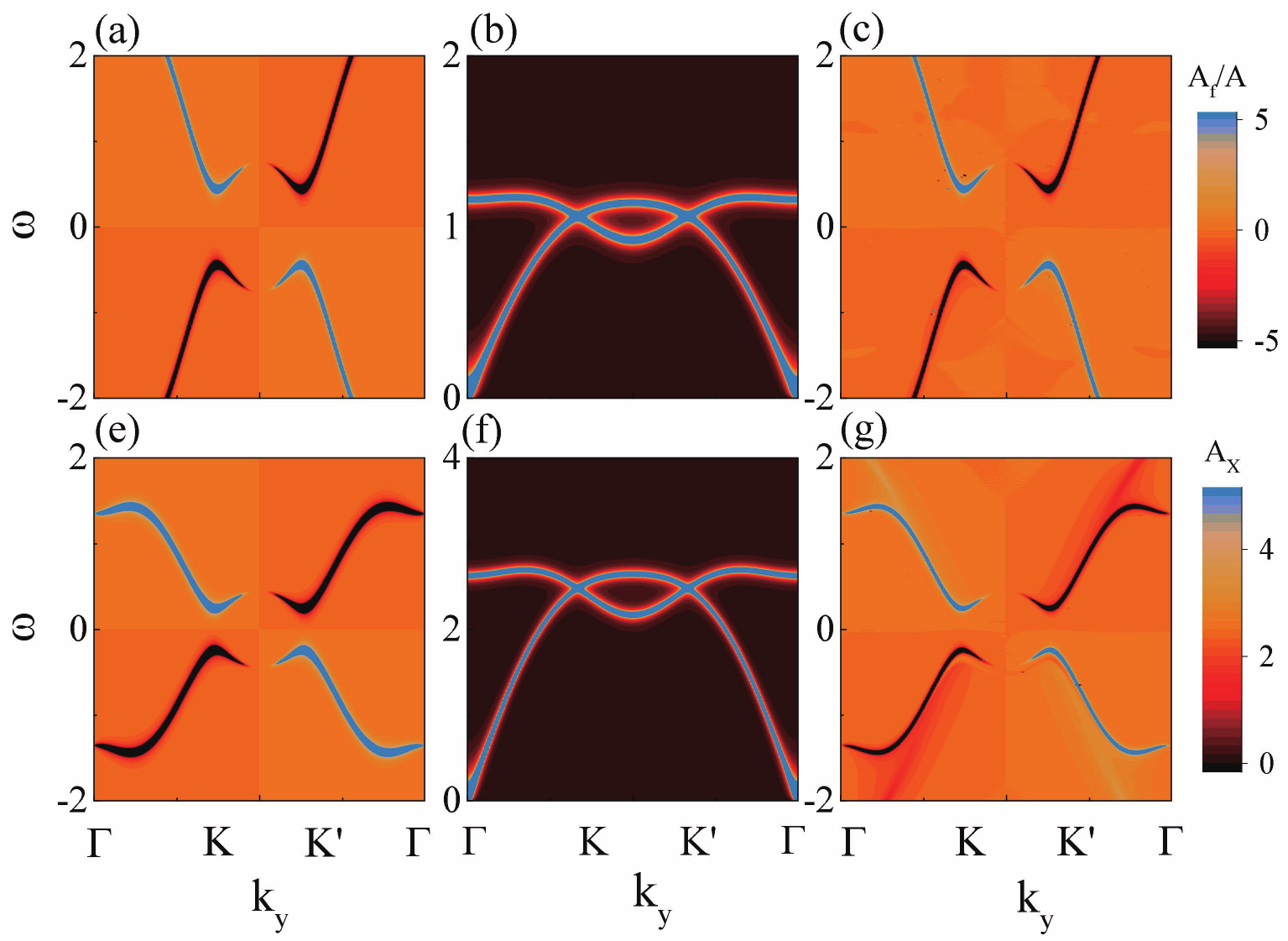}
\caption{(Color online) First row: (a)the bulk spectrum function of the fermionic spinon $A^{ss}_{f\uparrow}(\bm{k},\omega)-A^{ss}_{f\downarrow}(\bm{k},\omega)$, (b)bosonic chargon $A^{ss}_{X}(\bm{k},\omega)$,
and (c)electron $A^{ss}_{\uparrow}(\bm{k},\omega)-A^{ss}_{\downarrow}(\bm{k},\omega)$ on sublattice A along the zigzag direction in the chiral TBI state
at $\lambda=0.5$ and $U=0.5$; Second row: the same data at $\lambda=0.5$ and $U=2.7$.   \label{Bulk-Dispersion-Sym}}
\end{figure}
In the slave-rotor representation on the basis of charge-spin separation, an electron operator is decomposed into the direct product of a fermionic spinon operator $f_{i\sigma}$ and a constrained bosonic chargon operator
$X_{i}$ with $|X_{i}|^2=1$. As a result, within the saddle-point approximation, the electron Green's function is expressed as a convolution of the spinon and chargon Green's function[See Eq.\eqref{ELE-GF}],
thus it is necessary to
study the spectrum function for the spin and charge degree of freedom, as well as the electron spectral weight distribution in momentum space at the same time.
In Fig.\ref{Bulk-Dispersion-Sym}, we study the bulk spectrum function of (a)the fermionic spinon $A^{ss}_{f\uparrow}(\bm{k},\omega)-A^{ss}_{f\downarrow}(\bm{k},\omega)$, (b)the bosonic chargon $A^{ss}_{X}(\bm{k},\omega)$,
and (c)electron $A^{ss}_{\uparrow}(\bm{k},\omega)-A^{ss}_{\downarrow}(\bm{k},\omega)$ on sublattice A along the zigzag direction in the chiral TBI state at $\lambda=0.5$ with $U=0.5$ and 2.7 in the first and second
row, respectively, where the momenta $\Gamma$, $K$, and $K$' are illustrated in Fig. \ref{BZ} from Appendix \ref{Lattice setup}. We first note that the spectrum function on sublattice B can be reached by exchanging the energy dispersion within $0<k_y<\pi$ and $\pi<k_y<2\pi$ as a result of the chiral symmetry in the chiral TBI state.
The spinon, chargon, and electron spectrum function exhibit similar behaviors at small and moderate interaction strengths before the occurrence of metal-insulator Mott transition in the chiral TBI state, except two
important distinctions: (i)the bandwidth and energy gap in the spinon and electron energy spectrum are renormalized by the onsite electron coulomb interaction, especially for large strengths, and the diminishing energy gap
with the increasing interaction strength is consistent with the reduced effective SOC strength[See Fig. \ref{Paras-Chiral-Sym-NSym}(d)];
(ii)in contrast to the diminished bandwidth in the spinon and electron energy spectrum, the chargon bandwidth is significantly increased at strong interactions due to the increment of $\rho_{\rm A/B}$.
Moreover, the electron spectrum function exhibits similar behaviors to the spinon spectral weight distribution in momentum space
except that the electron quasiparticle spectral weight is reduced by the renormalization factor $Z_{ss'}$,
which is shown in Eq.\eqref{ELE-GF}. In addition, the chargon energy dispersion consists of two branches with the lower one responsible for the gapless bosonic excitations in the chiral TBI state,
and these two energy bands intersect with each other at the Dirac point $\bm{K}$ and $\bm{K}'$, which thus leads to zero energy gap between them.

\begin{figure}[h!]
\centering
\includegraphics[scale=0.3]{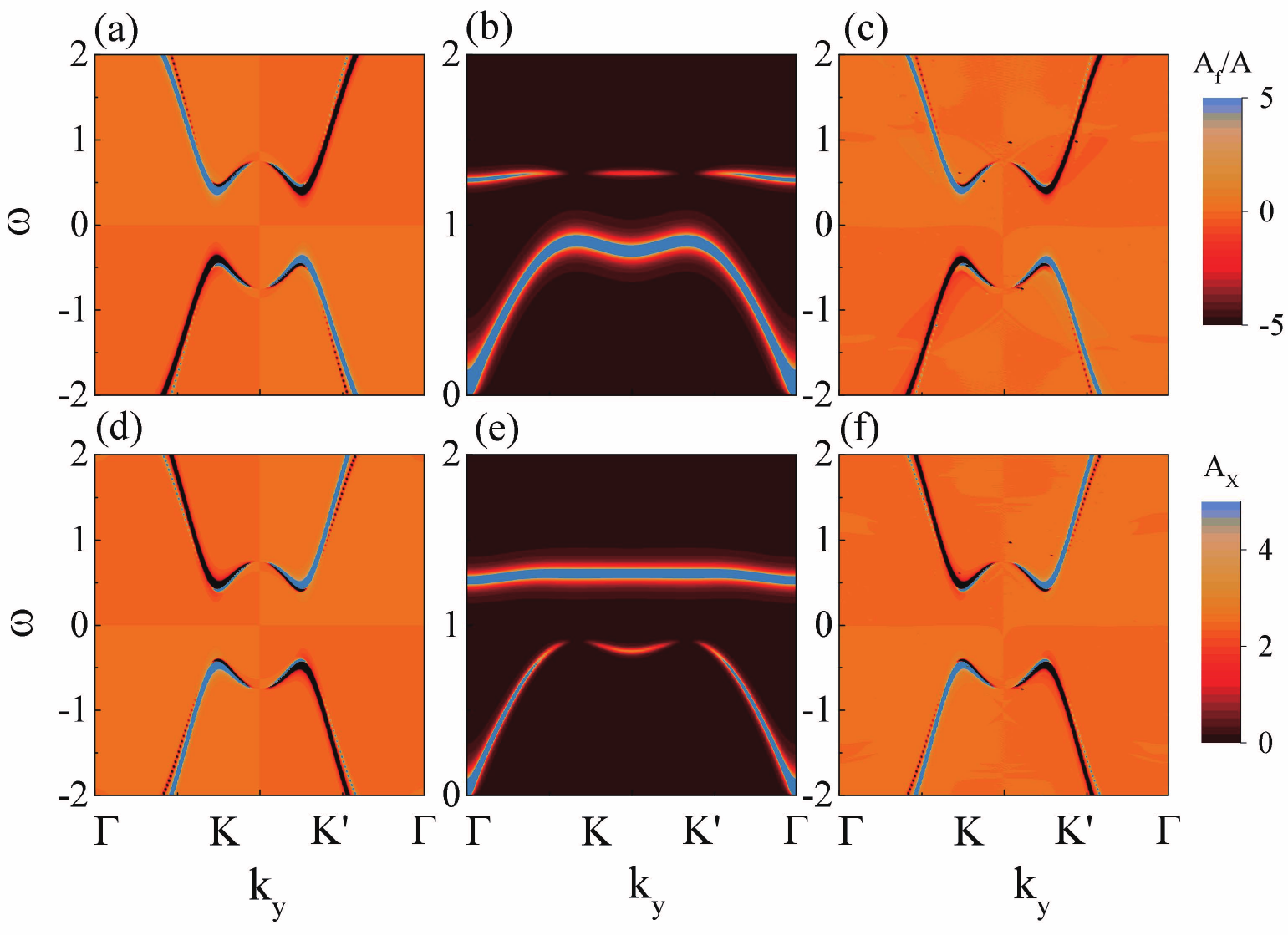}
\caption{(Color online) First row: (a)the spectrum function of the fermionic spinon $A^{ss}_{f\uparrow}(\bm{k},\omega)-A^{ss}_{f\downarrow}(\bm{k},\omega)$, (b)bosonic chargon $A^{ss}_{X}(\bm{k},\omega)$,
and (c)electron $A^{ss}_{\uparrow}(\bm{k},\omega)-A^{ss}_{\downarrow}(\bm{k},\omega)$ on sublattice A along the zigzag direction in the non-chiral TBI state
at $\lambda=0.5$ and $U=0.5$; Second row: the same data for sublattice B.   \label{Bulk-Dispersion-NSym05}}
\end{figure}
In the non-chiral TBI state, the spontaneously broken chiral symmetry in the presence of electron coulomb interaction destroys the relation between the spectrum function on sublattice A and B,
thus we in Fig.\ref{Bulk-Dispersion-NSym05} study the spectrum function of the fermionic spinon $A^{ss}_{f\uparrow}(\bm{k},\omega)-A^{ss}_{f\downarrow}(\bm{k},\omega)$, bosonic chargon $A^{ss}_{X}(\bm{k},\omega)$,
and electron $A^{ss}_{\uparrow}(\bm{k},\omega)-A^{ss}_{\downarrow}(\bm{k},\omega)$ along the zigzag direction in the non-chiral TBI state at $\lambda=0.5$ and $U=0.5$ on sublattice A(first row) and B(second row), respectively.
As shown in the first and third column of Fig.\ref{Bulk-Dispersion-NSym05}, apart from the leading energy band with the majority of spectral weight which appears in the chiral TBI state, there emerges the other minor energy
band in the vicinity of the leading band with much less quasiparticle spectral weight at small interaction strengths, which directly reflects the spontaneously broken chiral symmetry. Moreover, the gapless bosonic excitations
shown in the second column of Fig.\ref{Bulk-Dispersion-NSym05} validate our newly found non-chiral TBI state in the presence of intrinsic SOC and onsite electron coulomb interaction.
In addition, the two bosonic energy branches shown in Fig.\ref{Bulk-Dispersion-Sym}(b) intersecting with each other at
$\omega \approx 1t$ in the chiral TBI state with $U=0.5$ are now separated by an energy gap in the non-chiral TBI state, and whether the opening of this gap associates with a topological phase transition is under investigation.
The related results will be presented else where.

\begin{figure}[h!]
\centering
\includegraphics[scale=0.3]{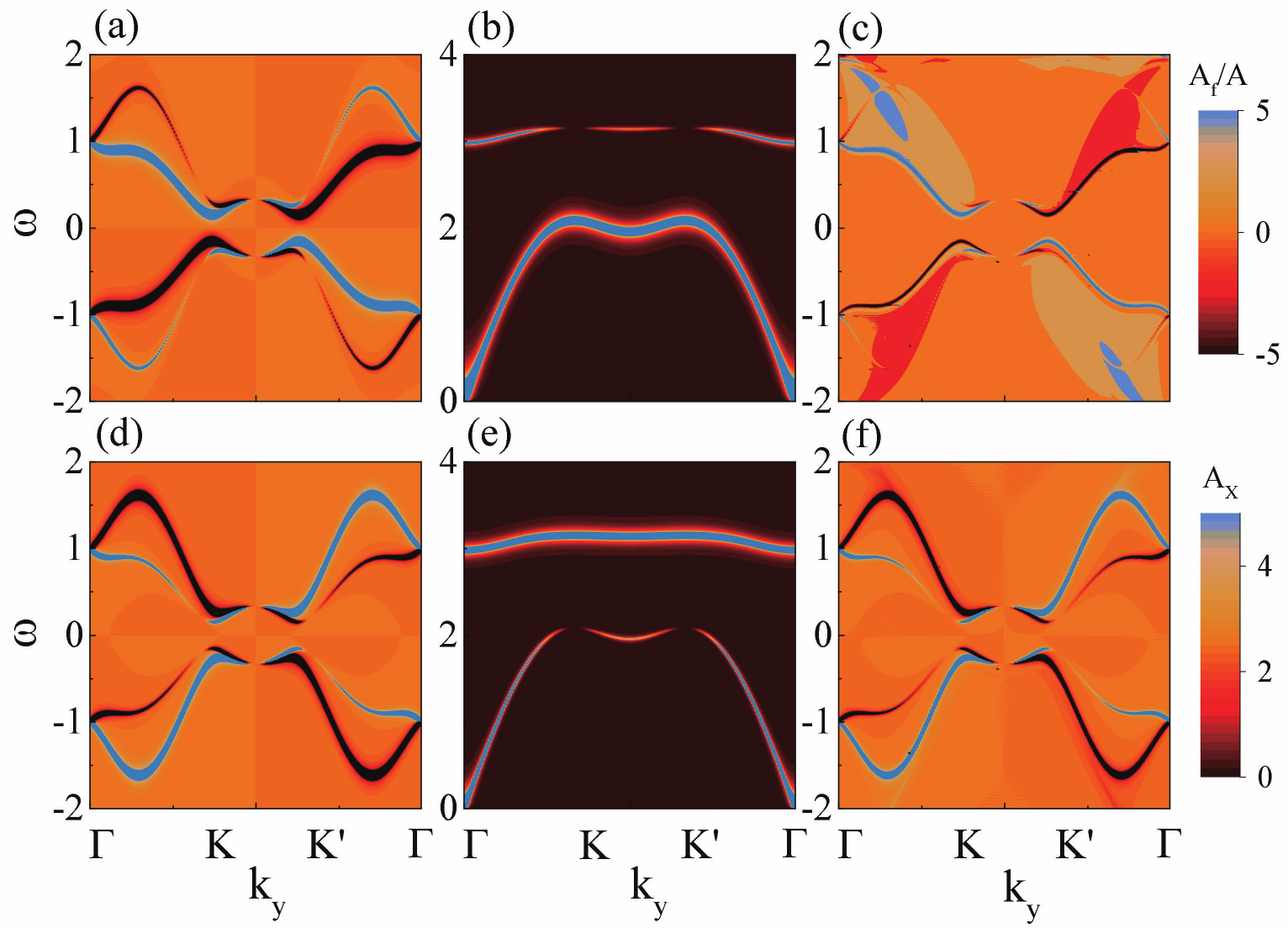}
\caption{(Color online) First row: (a)the spectrum function of the fermionic spinon $A^{ss}_{f\uparrow}(\bm{k},\omega)-A^{ss}_{f\downarrow}(\bm{k},\omega)$, (b)bosonic chargon $A^{ss}_{X}(\bm{k},\omega)$,
and (c)electron $A^{ss}_{\uparrow}(\bm{k},\omega)-A^{ss}_{\downarrow}(\bm{k},\omega)$ on sublattice A along the zigzag direction in the non-chiral TBI state
at $\lambda=0.5$ and $U=2.7$; Second row: the same data for sublattice B.  \label{Bulk-Dispersion-NSym27}}
\end{figure}
From Fig.\ref{Bulk-Dispersion-NSym05}, the intrinsic SOC together with the electron coulomb interaction spontaneously break the chiral symmetry of the KM model, which is manifested by the presence of a minor energy band
with low spectral weight apart from the leading energy band appearing in the chiral TBI state, and it would be interesting to study how this minor energy band evolves with the increasing interaction strength.
We in Fig.\ref{Bulk-Dispersion-NSym27} study the spectrum function $A^{ss}_{f\uparrow}(\bm{k},\omega)-A^{ss}_{f\downarrow}(\bm{k},\omega)$, $A^{ss}_{X}(\bm{k},\omega)$,
and $A^{ss}_{\uparrow}(\bm{k},\omega)-A^{ss}_{\downarrow}(\bm{k},\omega)$ along the zigzag direction in the non-chiral TBI state at $\lambda=0.5$ and $U=2.7$ on sublattice A(first row) and B(second row), respectively.
As shown in Fig.\ref{Bulk-Dispersion-NSym27}(a) and (d), the minor energy band signaling the spontaneously broken chiral symmetry becomes pronounced at $U=2.7$ with much higher spectral weight than
$U=0.5$, and the separation between the leading and minor energy band is significantly enlarged. Moreover, as shown in Fig.\ref{Bulk-Dispersion-NSym27}(c) and (f), the inequivalent renormalization factor $Z_{ss}$ on
sublattice A and B suppresses the electron quasiparticle spectral weight localized on two sublattices differently, which then leads to a long-range charge order with different electron occupation on sublattice A and B, respectively.
In addition, the bosonic chargon energy dispersion shown in the second column are similar to $U=0.5$ shown in Fig.\ref{Bulk-Dispersion-NSym05} except that the bandwidth is significantly increased as the chiral TBI state.

\subsection{The edge-state energy dispersion of the non-chiral TBI state and its real-space distribution}\label{Edge-Modes-EL}

In the interacting KM model, the spin-dependent topological invariant can be calculated via the single particle Green's function $G_{\sigma}(\bm{k},\omega)$ for electrons~\cite{Ishikawa86,Volovik03,Wang10,Blason23,Wagner24}, while for interaction strengths
smaller than its critical value $U_{\rm Mott}$, the single particle energy gap coming from the intrinsic SOC remains open when the on-site coulomb interaction is strengthened, which further indicates
that though the electron coulomb interaction leads to pronounced renormalization effects to the single-particle spectrum function, the topological invariant of this system remains unchanged~\cite{Wang10}.
Moreover, according to the usual bulk-boundary correspondence which is assumed to hold before the occurrence of the metal-insulator Mott transition, the physics of the edge modes for a nanoribbon directly reflects
the topology of the bulk system. Therefore, exploring how the electron coulomb interaction affects the edge mode of a cylindrical geometry periodic in the zigzag direction can not only unveil the interaction's effects on the edge states,
especially for the novel TBI state with spontaneously broken chiral symmetry induced by unequal $\rho_A$ and $\rho_B$ in the presence of onsite interaction and intrinsic SOC, but also further demonstrates that the topology
of the system keeps unchanged compared to the non-interacting KM model even in the above new TBI state because the energy gap coming from the intrinsic SOC keeps open as the increment of interaction strength for $U<U_{\rm Mott}$.

In Appendix \ref{Zigzag-Nanoribbon}, we have generalized the slave rotor method to a cylindrical geometry periodic in the zigzag direction, and the fermionic spinon and bosonic chargon Green's function have been
directly obtained from the Lagrange in the slave rotor representation similar to the bulk system with periodic boundary conditions, then the electron Green's function can be expressed as a convolution of the fermionic spinon
and bosonic chargon Green's function. We can then use the spectrum function $A_{\sigma}(k_y,\omega)$ and $A_{\sigma}(i_x,i_x,k_y,\omega)$ defined in Appendix \ref{Zigzag-Nanoribbon} to systematically study the
electron energy dispersion versus $k_y$ of the edge mode and its real-space distribution.

\begin{figure}[h!]
\centering
\includegraphics[scale=0.3]{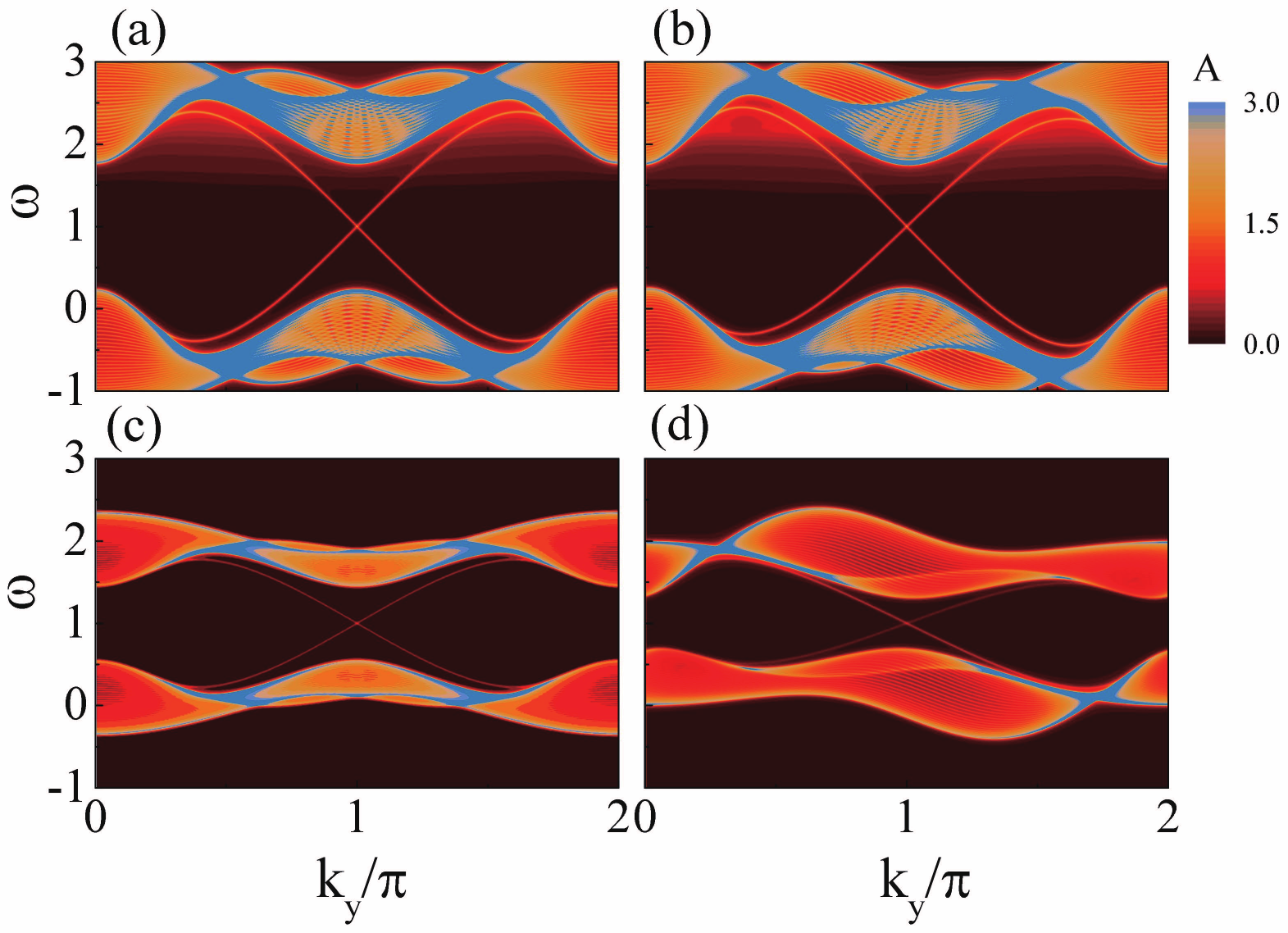}
\caption{(Color online) First row:(a) and (b) the spin-up electron spectrum function at $\lambda=0.5$ and $U=0.5$ of a nanoribbon with the periodic boundary condition along the zigzag direction
in the chiral and non-chiral TBI state, respectively; Second row:(c) and (d) the same data at $\lambda=0.5$ and $U=2.7$. \label{Edge-State}}
\end{figure}
We now study the edge-state energy dispersion at $\lambda=0.5$ and $U=$ 0.5, 2.7 with the periodic boundary condition along the zigzag direction in the chiral and non-chiral TBI state, respectively,
where the slave-rotor representation has been extended to calculate the spinon and chargon Green's function on the nanoribbon periodic in the zigzag direction
while with an open boundary condition in the armchair direction, then the electron edge-state spectrum function is obtained
via the convolution of the chargon and spinon Green's function within the Hartree-Fock approximation.
We note that the deviation of the edge-state energy band crossing point from zero energy shown in Fig.\ref{Edge-State} is due to the energy gap opened in the bosonic field $X$ on the
nanoribbon with a cylindrical geometry[c.f. Fig\ref{EDGE-U27-PSI-X} in Appendix \ref{Zigzag-Nanoribbon}] though its energy dispersion for the bulk system is gapless at $\bm{k}=\bm{0}$, while whether this energy gap can be eliminated by introducing the site- and bond-dependent self-consistent parameters
is still on investigation.
As shown in Fig.\ref{Edge-State}(a) and (c), in the chiral TBI state, there are two edge modes symmetric about $k_{y}=\pi$ with equal weight localized at opposite edges which propagate in the opposite direction,
indicating that the chiral symmetry of the KM model is maintained in the chiral TBI state under the influence of electron coulomb interaction.
In addition, the bandwidth of the edge-state energy dispersion and its spectral weight are significantly weakened
by strong interactions, which implies that the edge mode in this regime with $U\lesssim U_{\rm Mott}$ is difficult to observe experimentally. However, as shown in Fig.\ref{Edge-State}(b) and (d), in the non-chiral TBI state with
$\rho_{A}< \rho_{B}$, the energy dispersion of the bulk states and edge modes exhibit significant asymmetry with respect to $k_{y}=\pi$ especially for strong interactions.
More importantly, the gapless edge modes localized at opposite edges have different spectral weight and Fermi velocity which is more pronounced at large $U$'s.
The edge-state energy dispersion for spin-down electrons can be obtained by exchanging the spectral weight distribution between $0<k_y<\pi$ and $\pi<k_y<2\pi$ because of the time reversal symmetry.
In conclusion, the biggest difference between the chiral and non-chiral TBI state is that because of the spontaneously broken chiral symmetry in the latter state, there are inequivalent spin accumulation
at opposite edges of the cylinder periodic in the zigzag direction, which thus leads to a net spin current across the nanoribbon. In addition, the edge-mode spectral weight in the non-chiral TBI state is substantially larger
than the chiral TBI state at large interaction strengths which could make it observable experimentally. We emphasize that this new theoretical finding could give more inspirations to experimentalists
for designing novel topological spintronics devices.

\begin{figure}[h!]
\centering
\includegraphics[scale=0.3]{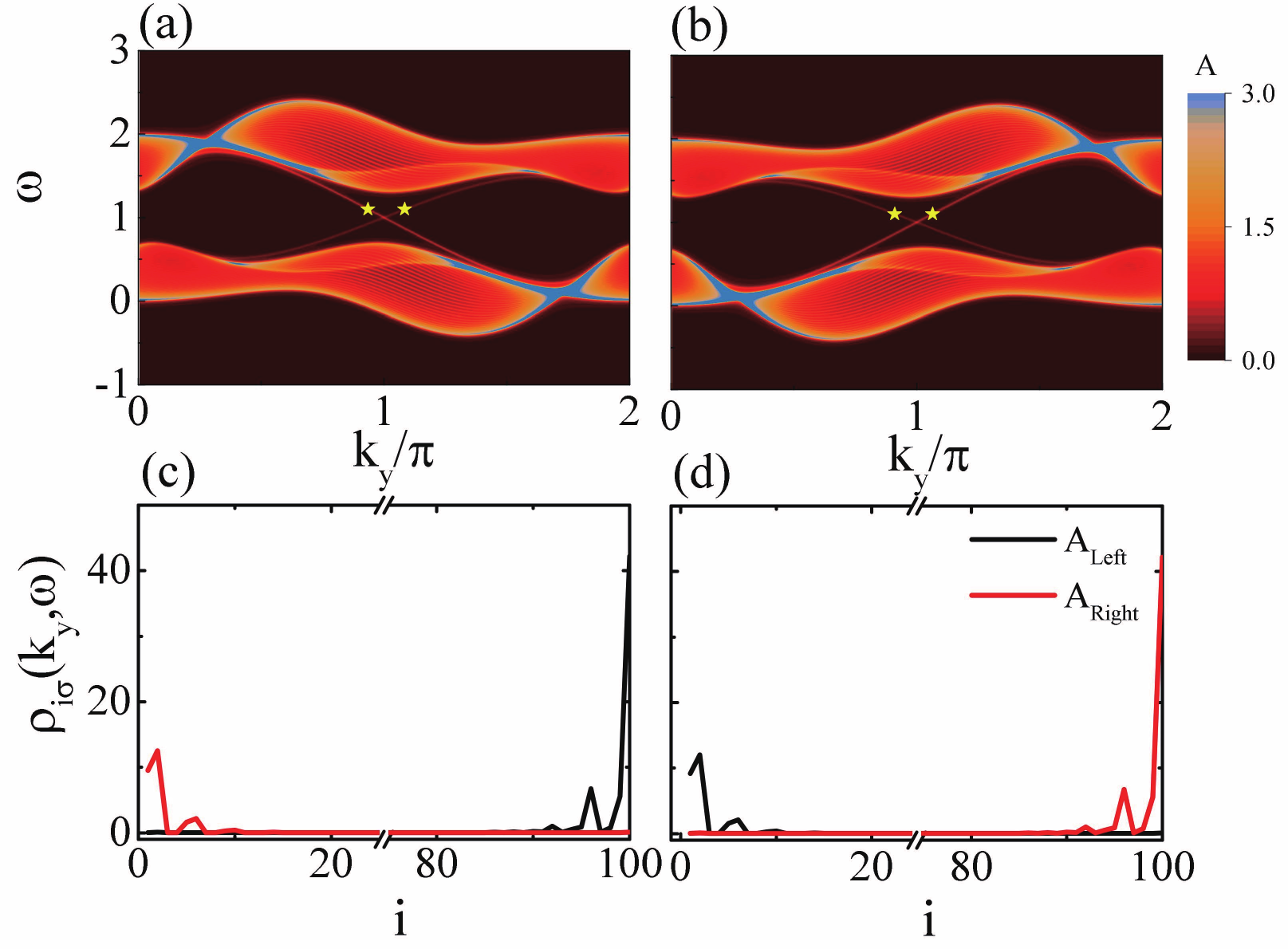}
\caption{(Color online) First row: the spectrum function of (a)up- and (b)down-spin electrons at $\lambda=0.5$ and $U=2.7$ for the nanoribbon with periodic boundary condition along the zigzag direction
in the non-chiral TBI state; Second row: the real-space spectral weight distribution $\rho_{i\sigma}(k_y,\omega)=A_{\sigma}(i_x,i_x,k_y,\omega)$ of four edge states marked by yellow pentagrams in the first row of this figure
for (c)up- and (d)down-spin electrons. The legends $A_{\rm left}$ and $A_{\rm right}$ represent the left and right edge states in the first row of this figure. \label{EDGE-U27SPINUP-DN}}
\end{figure}
To study the physics of the edge modes for up- and down-spin electrons more intuitively, we plot the edge-state spectral weight distribution as a function of $k_{y}$ and its real-space distribution
as a function of the atom location within a unit cell[See Fig.\ref{Cylinder} in Appendix \ref{Zigzag-Nanoribbon}] in the first and second row of Fig.\ref{EDGE-U27SPINUP-DN}, respectively, for up- and down-spin electrons at $\lambda=0.5$ and $U=2.7$.
As shown in Fig.\ref{EDGE-U27SPINUP-DN}(a) and (b), the energy dispersion for up- and down-spin electrons can be transformed into each other by exchanging spin and spectrum function within $0<k_y<\pi$ and $\pi<k_y<2\pi$ simultaneously
as a direct consequence of the time reversal symmetry, while for each spin the edge-state energy dispersion localized on one of the zigzag boundaries can not be obtained by exchanging the spectrum function within $0<k_y<\pi$
and $\pi<k_y<2\pi$ of the other boundary,
which indicates again the spontaneously broken chiral symmetry in the presence of electron coulomb interaction and intrinsic SOC. To further understand the physics of this special edge mode with spontaneously broken chiral symmetry, we investigate the real-space spectral weight distribution $\rho_{i\sigma}(k_y,\omega)=A_{\sigma}(i_x,i_x,k_y,\omega)$ for (c)up- and (d)down-spin electrons across the cylinder illustrated in Fig.\ref{Cylinder} with $k_{y}$ and $\omega$ marked by yellow pentagrams in the first row of Fig.\ref{EDGE-U27SPINUP-DN}. We find that the edge-state energy dispersion in Fig.\ref{EDGE-U27SPINUP-DN}(a) and (b) with more spectral weight are located around the lower boundary of the cylinder[See Fig.\ref{Cylinder}] consisting of B atoms for $\lambda>0$, while the edge-state modes with lower spectral weight are in the vicinity of the upper boundary which consists of A atoms. The inequivalent spectral weight of the edge modes on opposite boundaries gives rise to a net spin accumulation across this nanoribbon periodic in the zigzag direction while with an open boundary condition in the armchair direction. We emphasize that an opposite spin accumulation across this cylinder can be obtained for $\lambda<0$, and these two non-chiral TBI states with opposite spin current across the cylinder is separated by a topological phase transition at $\lambda=0$.

\section{Summary}\label{conclude}

We in this work use the slave rotor method to study the interacting Kane-Mele model, and find a new type of TBI state with spontaneously broken chiral symmetry which has lower energy compared to
the conventional TBI state maintaining the chiral symmetry in the presence of electron coulomb interaction, and thus this new TBI state is named by non-chiral TBI state. In addition, there exists a long range charge order with different electron
occupation on two sublattices in this non-chiral TBI state as well, which leads to a special helical edge state with inequivalent spin accumulation
at opposite edges of the cylinder with the periodic boundary condition in the zigzag direction, and then gives rise to a  net spin current across this system.
Moreover, this net spin accumulation coming from the spontaneously broken chiral symmetry
can be further strengthened by the nearest neighbor coulomb interaction between electrons which fosters the tendency towards a charge density wave state.

Recently, many efforts have been paid to the topological Mott insulator state at $U>U_{{\rm Mott}}$ with $U_{{\rm Mott}}$ being the critical interaction strength responsible for the metal-insulator Mott transition~\cite{Rachel10,Pesin10,Wagner24,Devakul22,Blason23,Guerci24}, in which some of them established a close relation between the dispersion of
electron Green's function zeros and spinon spectrum~\cite{Blason23,Wagner24}, or introduced an external sublattice field to produce band inversion then a topological phase transition
into the Chern insulator state~\cite{Devakul22}, as well as studied how the long-range magnetic order affects the physics in the topological insulator state and found the existence of the topological Kondo chiral semimetal
which can be transformed into the topological Kondo insulator state via applying a spatially random strain field~\cite{Guerci24}. However, up to now, we have not found any other work studying the charge order
state in the presence of electron coulomb interaction and the topological physics as a result of intrinsic SOC simultaneously, thus our work suggests a new research direction: how some kinds of long-range order associated
with spontaneously broken symmetries which is often caused by electron coulomb interaction affect the physics with respect to the usual topological transition and topological band insulator states.

\begin{acknowledgments}

This work was supported by the National Key Research and Development Program of
China under Grant Nos. 2023YFA1406500 and 2021YFA1401803, the National Natural
Science Foundation of  China (NSFC) under Grant Nos. 12222402, 92365101, 12474151,
12347101, 12247116, and 12274036, Beijing National Laboratory for Condensed Matter
Physics under Grant No. 2024BNLCMPKF025, the Fundamental Research Funds for the Central
Universities under Grant No. 2024IAIS-ZX002, the Chongqing Natural Science Foundation
under Grants Nos. CSTB2023NSCQ-JQX0024 and CSTB2022NSCQ-MSX0568, and the Special
Funding for Postdoctoral Research Projects in Chongqing under Grant No. 2024CQBSHTB3156.

\end{acknowledgments}

\appendix
\section{The lattice setup and its reciprocal lattice vectors}\label{Lattice setup}

In this work, the unit vectors in real space are chosen to be
\begin{equation}
\bm{a}_{1}=(\sqrt{3},1)\frac{1}{2} \;, \;\;\;\;\;\;\;\; \bm{a}_{2}=(-\sqrt{3},1)\frac{1}{2} \;,
\end{equation}
then three nearest-neighbor vectors $\bm{u}_{j=1,2,3}$ can be calculated via $\bm{u}_{j}=C_{3z}^{j-1}\bm{u}_1$ with $\bm{u}_1=(1,0)\frac{1}{\sqrt{3}}$,
while six next-nearest-neighbor vectors can be expressed as $\pm\bm{\gamma}_{j=1,2,3}$ with $\bm{\gamma}_{j}=C_{3z}^{j-1}\bm{\gamma}_1$ with $\bm{\gamma}_{1}=\bm{a}_{1}+\bm{a}_{2}$.

\begin{figure}[h!]
\centering
\includegraphics[scale=0.5]{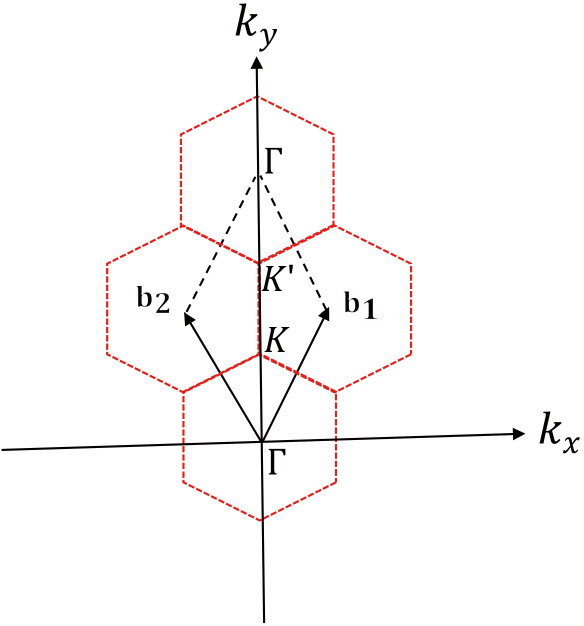}
\caption{(Color online) The schematic illustration of the Brillouin zone with the bases consisting of $\bm{b}_{1}$ and $\bm{b}_{2}$. \label{BZ}}
\end{figure}
Via $\bm{b}_{k}=\epsilon_{ijk}\frac{2\pi\bm{a}_{i}\times\bm{a}_{j}}{\bm{a}_{1}\cdot(\bm{a}_{2}\times\bm{a}_{3})}$ with $\epsilon_{ijk}$ being the Levi-Civita tensor,
the unit vectors in the reciprocal space are expressed as
\begin{equation}
\bm{b}_{1}=(\frac{1}{2},\frac{\sqrt{3}}{2})\frac{4\pi}{\sqrt{3}} \;, \;\;\;\;\;\;\;\; \bm{b}_{2}=(-\frac{1}{2},\frac{\sqrt{3}}{2})\frac{4\pi}{\sqrt{3}} \;,
\end{equation}
then the Brillouin zone is illustrated in Fig.\ref{BZ}. We introduce the following Fourier transform of hopping processes:
\begin{eqnarray}
V_{\bm{k}} &=& \sum_{\bm{u}}e^{i\bm{k}\cdot\bm{u}}\;,\;\;\;\;\epsilon_{\bm{k}s\sigma}=2\lambda\sum_{j=1,2,3}\cos(\bm{k}\cdot\bm{\gamma}_{j}\mp\sigma\pi/2)\;,\nonumber\\
\epsilon_{\bm{k}} &=& 2\lambda\sum_{j=1,2,3}\cos(\bm{k}\cdot\bm{\gamma}_{j})\;,
\end{eqnarray}
where the "$\mp$" correspond to sublattice A and B, respectively.

\begin{widetext}
\section{The derivation of the Green's function and the self-consistent equations}\label{GF-SC Eqs}

Following Refs.~\onlinecite{Florens02,Florens04,Rachel10,Wagner24}, the mean-field theory can be rigorously derived as the saddle point of the solvable non-linear sigma model in the large-$M$ limit with generalizing the $O(2)$ field $X_{is}(\tau)$ to $O(2M)$ field\cite{Cox93} $X_{is}^{\alpha}(\tau)$ supplemented to the local constraint $\sum_{\alpha}|X_{is}^{\alpha}|^2=M$, where $i$ and $s$ are the index of  lattice unit cell and sublattice, respectively. Then the Lagrange of the interacting KM model in the large $M$ limit can be expressed as by using the Hubbard-Stratonovich transformation
\begin{eqnarray}
L&=&\sum_{s=A,B}\sum_{i\alpha}\big[\frac{|\partial_{\tau}X_{is}^{\alpha}|^2-{h_{is}}^2}{2U}+h_{is}
+ \frac{h_{is}}{2U}(\partial_{\tau}X_{is}^{\alpha}{X_{is}^{\alpha}}^\ast-\partial_{\tau}{X_{is}^{\alpha}}^\ast X_{is}^{\alpha})\big]+ \sum_{is}\rho_{is}(\sum_{\alpha}|X_{is}^{\alpha}|^2-M)\nonumber\\
&+& \sum_{is}f_{is}^{\dagger}(\partial_{\tau}-h_{is})f_{is}+t\sum_{\langle ij\rangle}Q_{AB}^{X}(i,j)Q_{BA}^{f}(j,i)+\mathrm{H.C.} \nonumber\\
&-& t\sum_{\langle ij\rangle}\big[Q_{AB}^{X}(i,j)\sum_{\alpha}X_{iA}^{\alpha}{X_{jB}^{\alpha}}^{\ast}
+f_{iA}^{\dagger}f_{jB}Q_{BA}^{f}(j,i) +\mathrm{H.C.}\big]
-\lambda\sum_{s=A,B}\sum_{\langle\langle ij\rangle\rangle}Q_{ss}^{X}(i,j)Q_{ss}^{f}(j,i) \nonumber\\
&+& \lambda\sum_{s=A,B}\sum_{\langle\langle ij\rangle\rangle}\big[ Q_{ss}^{f}(j,i)f_{is}^{\dagger}
e^{i\tfrac{\pi}{2}\nu_{ij}\sigma^{z}}f_{js} + Q_{ss}^{X}(i,j)\sum_{\alpha}X_{is}^{\alpha}{X_{js}^{\alpha}}^{\ast} \big]
\end{eqnarray}
 with two set of complex field $Q_{ss'}^{X}(i,j)$ and $Q_{s's}^{f}(j,i)$.
In the limit $N_{s}$, $M\to \infty$ with fixed $N_{s}/M$ where $N_{s}$ represents the spin degeneracy, the solution of the above Lagrange can be obtained in the saddle-point approximation, and the translation invariance
as well as the point group symmetry of the system lead to the following Lagrange
\begin{eqnarray}\label{Lagrange-LargeM}
L&=&\sum_{s=A,B}\sum_{i}\big[\frac{|\partial_{\tau}X_{is}|^2-{h_{s}}^2}{U}+h_{s}
+ \frac{h_{s}}{U}(\partial_{\tau}X_{is}X_{is}^\ast-\partial_{\tau}X_{is}^\ast X_{is})\big]+ \sum_{is}\rho_{s}(|X_{is}|^2-1)\nonumber\\
&+& \sum_{is}f_{is}^{\dagger}(\partial_{\tau}-h_{s})f_{is}+3Nt[Q_{AB}^{X}Q_{BA}^{f}+\mathrm{H.C.}]
- t\sum_{\langle ij\rangle}\big[Q_{AB}^{X}X_{iA}X_{jB}^{\ast} + f_{iA}^{\dagger}f_{jB}Q_{BA}^{f} + \mathrm{H.C.}\big]\nonumber\\
&-& 6N\lambda\sum_{s=A,B}Q_{ss}^{X}Q_{ss}^{f}
+ \lambda\sum_{s=A,B}\sum_{\langle\langle ij\rangle\rangle}\big[ Q_{ss}^{f}f_{is}^{\dagger}e^{i\tfrac{\pi}{2}\nu_{ij}\sigma^{z}}f_{js} + Q_{ss}^{X}X_{is}X_{js}^{\ast} \big]\;,
\end{eqnarray}
where the interaction strength $U$ has been rescaled to its half to ensure the right gap behaviors~\cite{Florens02,Florens04}. The self-consistent equations are determined by using the least action principle:
\begin{subequations}\label{Self-ConstEq}
\begin{eqnarray}
Q_{ss}^{f}&=&Z_{ss}+\frac{T}{N6\lambda}\sum_{\bm{k}i\nu}\epsilon_{\bm{k}}G_{X}^{ss}(\bm{k},i\nu)\;,\\
Q_{BA}^{f}&=&Z_{BA}+ \frac{T}{N3}\sum_{\bm{k}i\nu}V_{\bm{k}}^{\ast}G_{X}^{AB}(\bm{k},i\nu)\;, \\
Q_{ss}^{X}&=&\frac{T}{6N\lambda}\sum_{\bm{k}i\omega\sigma} \epsilon_{\bm{k}s\sigma}G_{f\sigma}^{ss}(\bm{k},i\omega)e^{-i\omega 0^-}\;, \\
Q_{AB}^{X}&=& \frac{T}{N3}\sum_{\bm{k}i\omega\sigma}V_{\bm{k}}G_{f\sigma}^{BA}(\bm{k},i\omega)e^{-i\omega0^-}\;,\\
1&=&Z_{ss}+\frac{T}{N}\sum_{\bm{k}\neq0i\nu}G_{X}^{ss}(\bm{k},i\nu)\;, \\
\frac{T}{N}\sum_{\bm{k}i\omega\sigma}G_{f\sigma}^{ss}(\bm{k},i\omega)e^{-i\omega0^{-}}&=&1-\frac{2h_{s}}{U}
- \frac{T}{UN}\sum_{\bm{k}i\nu}i\nu G_{X}^{ss}(\bm{k},i\nu)[e^{-i\nu0^{-}}+e^{-i\nu0^{+}}] \;,
\end{eqnarray}
\end{subequations}
where $\beta$ and $N$ are the inverse of temperature $T$ and lattice size, respectively; $i\omega=i(2n+1)\pi/\beta$ and $i\nu=i2n\pi/\beta$ are the fermionic and bosonic Matsubara frequency, respectively;
the quasiparticle renormalization factor $Z_{ss'}$ are determined by the condensation of the bosonic field $X$,
\begin{equation}
Z_{ss'}=\frac{1}{N}\langle {X_{\bm{k}=0}^{s}}^{\ast}(\tau)X_{\bm{k}=0}^{s'}(\tau)\rangle=\sqrt{Z_{ss}Z_{s's'}}\;.
\end{equation}
The last two sets of equations in Eq.\eqref{Self-ConstEq} are used to self-consistently determine the Lagrange multipliers $\rho_{s}$ and $h_{s}$, respectively. Following the same procedure as the previous work by
N. Wagner, D. Guerci, and et al.\cite{Wagner24}, the components of the Green's function $G_{f\sigma}^{ss'}(i-j,\tau)=-\langle T_{\tau}f_{is\sigma}(\tau) f_{js'\sigma}^{\dagger}(0)\rangle$ and $G_{X}^{s's}(j-i,0-\tau)=\langle T_{\tau}X_{js'}(0)X_{is}^{\ast}(\tau)\rangle$ in momentum-energy space can be obtained via the Lagrange \eqref{Lagrange-LargeM} as
\begin{subequations}\label{GF-Psi-X-1}
\begin{eqnarray}
G_{f\sigma}(\bm{k},i\omega)&=&\frac{\left[
\begin{array}{ll}
i\omega+h_{B}-Q_{BB}^{f}\epsilon_{\bm{k}B\sigma}& -tQ_{BA}^{f}V_{\bm{k}}\\
-tQ_{AB}^{f}V_{\bm{k}}^{\ast}& i\omega+h_{A}-Q_{AA}^{f}\epsilon_{\bm{k}A\sigma} \end{array}   \right]}
{[i\omega+h_{A}-Q_{AA}^{f}\epsilon_{\bm{k}A\sigma}][i\omega+h_{B}-Q_{BB}^{f}\epsilon_{\bm{k}B\sigma}]-t^2|Q_{AB}^{f}V_{\bm{k}}|^2}\;, \\
G_{X}(\bm{k},i\nu)&=& \frac{-\left[
\begin{array}{ll}
\frac{(i\nu)^2}{U}+\frac{2h_{B}i\nu}{U}-\rho_{B}-Q_{BB}^{X}\epsilon_{\bm{k}}& -tQ_{BA}^{X}V_{\bm{k}}\\
-tQ_{AB}^{X}V_{\bm{k}}^{\ast}& \frac{(i\nu)^2}{U}+\frac{2h_{A}i\nu}{U}-\rho_{A}-Q_{AA}^{X}\epsilon_{\bm{k}}\end{array}   \right]}
{[\frac{(i\nu)^2}{U}+\frac{2h_{A}i\nu}{U}-\rho_{A}-Q_{AA}^{X}\epsilon_{\bm{k}}][\frac{(i\nu)^2}{U}+\frac{2h_{B}i\nu}{U}-\rho_{B}-Q_{BB}^{X}\epsilon_{\bm{k}}]-|Q_{AB}^{X}V_{\bm{k}}|^2} \;.
\end{eqnarray}
\end{subequations}
Within the saddle-point approximation, the interacting electron Green's function can be approximated as
$G_{\sigma}^{ss'}(i-j,\tau)\approx-\langle T_{\tau}f_{is\sigma}(\tau) f_{js'\sigma}^{\dagger}(0)\rangle
\langle T_{\tau}X_{js'}(0)X_{is}^{\ast}(\tau)\rangle=G_{f\sigma}^{ss'}(i-j,\tau)G_{X}^{s's}(j-i,0-\tau)$,
where $G_{f\sigma}^{ss'}(i-j,\tau)$ and $G_{X}^{s's}(j-i,-\tau)$ are the real-space Green's function of the spin and charge degree of freedom, respectively.
Then by taking the contribution from the bosonic condensation of $X$ into account~\cite{Rachel10,Wagner24}, the electron Green's function is expressed as
\begin{eqnarray}\label{ELE-GF}
G_{\sigma}^{ss'}(\bm{k},i\omega) \approx Z_{ss'}G_{f\sigma}^{ss'}(\bm{k},i\omega)
+ \frac{T}{N}\sum_{\bm{q}i\nu}G_{f\sigma}^{ss'}(\bm{k}+\bm{q},i\omega+i\nu)G_{X}^{s's}(\bm{q},i\nu)\;,
\end{eqnarray}
thus the electron Green's function can be calculated on the basis of the Green's function in Eq.\eqref{GF-Psi-X-1}.

In the half-filled system, the particle-hole symmetry requires that $h_{s}=\mu=0$ and the last set of equations in Eq.\eqref{Self-ConstEq} are fulfilled automatically, then the Green's functions in Eq.\eqref{GF-Psi-X-1}
can be rewritten by factorizing their denominators as
\begin{subequations}\label{GF-Psi-X-2}
\begin{eqnarray}
G_{f\sigma}(\bm{k},i\omega)&=&\frac{\left[
\begin{array}{ll}
i\omega-Q_{BB}^{f}\epsilon_{\bm{k}B\sigma}& -tQ_{BA}^{f}V_{\bm{k}}\\
-tQ_{AB}^{f}V_{\bm{k}}^{\ast}& i\omega -Q_{AA}^{f}\epsilon_{\bm{k}A\sigma} \end{array}   \right]}
{[i\omega-\epsilon_{+\bm{k}\sigma}][i\omega-\epsilon_{-\bm{k}\sigma}]}\;,\\
G_{X}(\bm{k},i\nu)&=&\frac{-U\left[
\begin{array}{ll}
(i\nu)^2-U\rho_{B}-UQ_{BB}^{X}\epsilon_{\bm{k}}& -tUQ_{BA}^{X}V_{\bm{k}}\\
-tUQ_{AB}^{X}V_{\bm{k}}^{\ast}& (i\nu)^2-U\rho_{A}-UQ_{AA}^{X}\epsilon_{\bm{k}}\end{array}   \right]}
{[i\nu-\omega_{1}(\bm{k})][i\nu-\omega_{2}(\bm{k})][i\nu-\omega_{3}(\bm{k})][i\nu-\omega_{4}(\bm{k})]} \;,
\end{eqnarray}
\end{subequations}
where
\begin{eqnarray*}
\epsilon_{\pm\bm{k}\sigma}&=&\frac{1}{2}\Big\{ Q_{AA}^{f}\epsilon_{\bm{k}A\sigma}+Q_{BB}^{f}\epsilon_{\bm{k}B\sigma}\pm
\sqrt{[ -Q_{AA}^{f}\epsilon_{\bm{k}A\sigma}+Q_{BB}^{f}\epsilon_{\bm{k}B\sigma}]^2+4t^2|Q_{AB}^{f}V_{\bm{k}}|^2} \Big\} \;,\\
\omega_{1}(\bm{k})&=&\sqrt{\frac{U}{2}}\Big\{\rho_{A}+Q_{AA}^{X}\epsilon_{\bm{k}}+\rho_{B}+Q_{BB}^{X}\epsilon_{\bm{k}} +
\sqrt{[\rho_{A}+Q_{AA}^{X}\epsilon_{\bm{k}}-\rho_{B}-Q_{BB}^{X}\epsilon_{\bm{k}}]^2+4|Q_{AB}^{X}V_{\bm{k}}|^2 }   \Big\}^{1/2}\;, \\
\omega_{2}(\bm{k})&=&-\sqrt{\frac{U}{2}}\Big\{\rho_{A}+Q_{AA}^{X}\epsilon_{\bm{k}}+\rho_{B}+Q_{BB}^{X}\epsilon_{\bm{k}} +
\sqrt{[\rho_{A}+Q_{AA}^{X}\epsilon_{\bm{k}}-\rho_{B}-Q_{BB}^{X}\epsilon_{\bm{k}}]^2+4|Q_{AB}^{X}V_{\bm{k}}|^2 }   \Big\}^{1/2}\;, \\
\omega_{3}(\bm{k})&=&\sqrt{\frac{U}{2}}\Big\{\rho_{A}+Q_{AA}^{X}\epsilon_{\bm{k}}+\rho_{B}+Q_{BB}^{X}\epsilon_{\bm{k}} -
\sqrt{[\rho_{A}+Q_{AA}^{X}\epsilon_{\bm{k}}-\rho_{B}-Q_{BB}^{X}\epsilon_{\bm{k}}]^2+4|Q_{AB}^{X}V_{\bm{k}}|^2 }   \Big\}^{1/2}\;, \\
\omega_{4}(\bm{k})&=&-\sqrt{\frac{U}{2}}\Big\{\rho_{A}+Q_{AA}^{X}\epsilon_{\bm{k}}+\rho_{B}+Q_{BB}^{X}\epsilon_{\bm{k}} -
\sqrt{[\rho_{A}+Q_{AA}^{X}\epsilon_{\bm{k}}-\rho_{B}-Q_{BB}^{X}\epsilon_{\bm{k}}]^2+4|Q_{AB}^{X}V_{\bm{k}}|^2 }   \Big\}^{1/2}\;.
\end{eqnarray*}
We note that the nonzero Lagrange multiplier $h_{s}$ at finite hole doping concentrations changes the denominator of $G_{X}(\bm{k},i\nu)$ into a general 4-order equation for the bosonic excitation dispersion,
which makes the expression of $\omega_{1,\cdots,4}(\bm{k})$ very complicated, and thus are not listed here.
In a topological band insulator with nonzero quasiparticle renormalization factor $Z_{ss'}$, the condensation of the bosonic field $X$ is required which then leads to the gapless
bosonic excitations at $\bm{k}=\bm{0}$, i.e.,
\begin{eqnarray}
 \Big\{ \rho_{A}+Q_{AA}^{X}\epsilon_{\bm{k}}+\rho_{B}+Q_{BB}^{X}\epsilon_{\bm{k}}
\pm \sqrt{[\rho_{A}+Q_{AA}^{X}\epsilon_{\bm{k}}-\rho_{B}-Q_{BB}^{X}\epsilon_{\bm{k}}]^2+4|Q_{AB}^{X}V_{\bm{k}}|^2 } \Big\}_{\bm{k}=\bm{0}}=0\;, \nonumber\\
\end{eqnarray}
indicating that
\begin{eqnarray}
\big[ \rho_{A}+\rho_{B}+6\lambda(Q_{AA}^{X}+Q_{BB}^{X}) \big]^2
= [\rho_{A}-\rho_{B}+6\lambda(Q_{AA}^{X}-Q_{BB}^{X})]^2+4|Q_{AB}^{X}D|^2 \;,
\end{eqnarray}
then
\begin{eqnarray}
\rho_{B} = \frac{|Q_{AB}^{X}D|^2}{\rho_{A}+6\lambda Q_{AA}^{X}}-6\lambda Q_{BB}^{X}
\end{eqnarray}
with $D=3$ being the half bandwidth of the tight-binding model on a hexagon lattice. Obviously, in the topological band insulator state, the condensed bosonic field $X$ leads to a relation between the Lagrange multipliers
$\rho_{A}$ and $\rho_{B}$, which then gives rise to a redundancy for the determination of $\rho_{A}$. Moreover, if $\rho_{A}$ is taken to be the value in the chiral TBI state, i.e.,
$\rho_{A}=Q_{AB}^{X}D-6\lambda Q_{AA}^{X}$, $\rho_{B}$ will be reduced as $\rho_{B}=Q_{AB}^{X}D-6\lambda Q_{BB}^{X}$, then with $Q_{AA}^{X}=Q_{BB}^{X}$, we have $\rho_{A}=\rho_{B}=Q_{AB}^{X}D-6\lambda Q_{ss}^{X}$.
The Lagrange multiplier $\rho_{A}$ can be generally expressed as
\begin{equation}
\rho_{A} = Q_{AB}^{X}D + f(\lambda,U)
\end{equation}
with $ f(\lambda\to 0,U)=0$ such that the solution of the Hubbard model can be reached at zero spin-orbit coupling strength. We have chosen different $f$ function such as $f(\lambda,U)=0, -\alpha \lambda/U$,
and found that these $f$ functions do not qualitatively alter the mean-field parameters in Eq.\eqref{Self-ConstEq}, thus in the following calculations we choose $\rho_{A} = Q_{AB}^{X}D$, then we have
\begin{equation}
\rho_{B} = \rho_{A}-C\frac{2\rho_{A}+C}{\rho_{A}+C}
\end{equation}
with $C=6\lambda Q_{AA}^{X}$, which clearly demonstrates that in the presence of onsite coulomb interaction between electrons, the nonzero SOC strength $\lambda$ gives rise to inequivalent Lagrange multipliers $\rho_{A}$ and $\rho_{B}$.
Furthermore, the inequivalent $\rho_{A}$ and $\rho_{B}$ lead to different renormalization factor $Z_{ss}$ on sublattice A and B, respectively,
which then gives rise to different electron occupation on two sublattices, i.e., the simplest form of charge order.
Moreover, in this charge order state the chiral symmetry of the non-interacting KM model is spontaneously broken in the presence of intrinsic SOC and electron coulomb interaction,
therefore this charge ordering TBI state with $\rho_{A}\neq\rho_{B}$ is named by the non-chiral TBI state.
Accordingly the topological insulator state with chiral symmetry is renamed by the chiral TBI state to better
demonstrate the difference between the topological band insulator state with $\rho_{A}=\rho_{B}$ and $\rho_{A}\neq \rho_{B}$, respectively.

Substituting Eq.\eqref{GF-Psi-X-2} into Eq.\eqref{Self-ConstEq}, the first six self-consistent equations can be expressed as
\begin{subequations}\label{Self-Eq-1}
\begin{eqnarray}
Q_{ss}^{f}&=&Z_{ss}+\frac{U}{N6\lambda}\sum_{\bm{k}\neq\bm{0}}\epsilon_{\bm{k}}\bigg\{ \frac{n_{\mathrm{B}}[\omega_{1}(\bm{k})]
\big[\omega_{1}(\bm{k})^2-U(\rho_{\bar{s}}+Q_{\bar{s}\bar{s}}^{X}\epsilon_{\bm{k}})\big] }
{\prod_{j\neq1}[\omega_{1}(\bm{k})-\omega_{j}(\bm{k})]} + \frac{n_{\mathrm{B}}[\omega_{2}(\bm{k})]
\big[\omega_{2}(\bm{k})^2-U(\rho_{\bar{s}}+Q_{\bar{s}\bar{s}}^{X}\epsilon_{\bm{k}})\big] }{\prod_{j\neq2}[\omega_{2}(\bm{k})-\omega_{j}(\bm{k})]} \nonumber\\
&+&\frac{n_{\mathrm{B}}[\omega_{3}(\bm{k})]
\big[\omega_{3}(\bm{k})^2-U(\rho_{\bar{s}}+Q_{\bar{s}\bar{s}}^{X}\epsilon_{\bm{k}})\big] }
{\prod_{j\neq3}[\omega_{3}(\bm{k})-\omega_{j}(\bm{k})]}
+\frac{n_{\mathrm{B}}[\omega_{4}(\bm{k})]
\big[\omega_{4}(\bm{k})^2-U(\rho_{\bar{s}}+Q_{\bar{s}\bar{s}}^{X}\epsilon_{\bm{k}})\big] }
{\prod_{j\neq4}[\omega_{4}(\bm{k})-\omega_{j}(\bm{k})]}\bigg\}\;,\\
Q_{ss}^{X}&=&\frac{1}{6N\lambda}\sum_{\bm{k}\sigma} \epsilon_{\bm{k}s\sigma}\big[
\frac{n_{\mathrm{F}}(\epsilon_{+\bm{k}\sigma})}{\epsilon_{+\bm{k}\sigma}-\epsilon_{-\bm{k}\sigma}}(\epsilon_{+\bm{k}\sigma}-Q_{\bar{s}\bar{s}}^{f}\epsilon_{\bm{k}\bar{s}\sigma})
-\frac{n_{\mathrm{F}}(\epsilon_{-\bm{k}\sigma})}{\epsilon_{+\bm{k}\sigma}-\epsilon_{-\bm{k}\sigma}}(\epsilon_{-\bm{k}\sigma}-Q_{\bar{s}\bar{s}}^{f}\epsilon_{\bm{k}\bar{s}\sigma})
\big]\;,\\
Q_{BA}^{f}&=& Z_{BA}-\frac{tU^2Q_{BA}^{X}}{N3}\sum_{\bm{k}\neq\bm{0}}|V_{\bm{k}}|^2 \bigg\{ \frac{n_{\mathrm{B}}[\omega_{1}(\bm{k})] }
{\prod_{j\neq1}[\omega_{1}(\bm{k})-\omega_{j}(\bm{k})]}
+\frac{n_{\mathrm{B}}[\omega_{2}(\bm{k})] }
{\prod_{j\neq2}[\omega_{2}(\bm{k})-\omega_{j}(\bm{k})]} \nonumber\\
&+&\frac{n_{\mathrm{B}}[\omega_{3}(\bm{k})] }
{\prod_{j\neq3}[\omega_{3}(\bm{k})-\omega_{j}(\bm{k})]}
+\frac{n_{\mathrm{B}}[\omega_{4}(\bm{k})] }
{\prod_{j\neq4}[\omega_{4}(\bm{k})-\omega_{j}(\bm{k})]}
\bigg\}\;, \\
Q_{AB}^{X}&=& \frac{-tQ_{AB}^{f}}{N3}\sum_{\bm{k}\sigma}\frac{|V_{\bm{k}}|^2}{\epsilon_{+\bm{k}\sigma}-\epsilon_{-\bm{k}\sigma}}
\big[ n_{\mathrm{F}}(\epsilon_{+\bm{k}\sigma})-n_{\mathrm{F}}(\epsilon_{-\bm{k}\sigma}) \big]\;,
\end{eqnarray}
where $\bar{s}$ represents the other sublattice for $s=$ A and B, and the quantity $Q_{AA/BB}^{X}$ at zero temperature can be simplified as
\begin{equation}
Q_{AA}^{X}=Q_{BB}^{X}=\frac{-1}{6N\lambda}\sum_{\bm{k}}\frac{\epsilon_{\bm{k}A\uparrow}^2(Q_{AA}^{f}+Q_{BB}^{f})}
{\sqrt{(Q_{AA}^{f}+Q_{BB}^{f})^2\epsilon_{\bm{k}A\uparrow}^2+4t^2|Q_{AB}^{f}V_{\bm{k}}|^2}} \;. \nonumber
\end{equation}
In addition, the self-consistent equations with respect to the Lagrange multipliers $\rho_{s}$ and $h_{s}$ read
\begin{eqnarray}
1&=&Z_{ss}+\frac{U}{N}\sum_{\bm{k}\neq0}\bigg\{ \frac{n_{\mathrm{B}}[\omega_{1}(\bm{k})]
\big[\omega_{1}(\bm{k})^2-U(\rho_{\bar{s}}+Q_{\bar{s}\bar{s}}^{X}\epsilon_{\bm{k}})\big] }
{\prod_{j\neq1}[\omega_{1}(\bm{k})-\omega_{j}(\bm{k})]}
+\frac{n_{\mathrm{B}}[\omega_{2}(\bm{k})]
\big[\omega_{2}(\bm{k})^2-U(\rho_{\bar{s}}+Q_{\bar{s}\bar{s}}^{X}\epsilon_{\bm{k}})\big] }
{\prod_{j\neq2}[\omega_{2}(\bm{k})-\omega_{j}(\bm{k})]} \nonumber\\
&+&\frac{n_{\mathrm{B}}[\omega_{3}(\bm{k})]
\big[\omega_{3}(\bm{k})^2-U(\rho_{\bar{s}}+Q_{\bar{s}\bar{s}}^{X}\epsilon_{\bm{k}})\big] }
{\prod_{j\neq3}[\omega_{3}(\bm{k})-\omega_{j}(\bm{k})]}
+\frac{n_{\mathrm{B}}[\omega_{4}(\bm{k})]
\big[\omega_{4}(\bm{k})^2-U(\rho_{\bar{s}}+Q_{\bar{s}\bar{s}}^{X}\epsilon_{\bm{k}})\big] }
{\prod_{j\neq4}[\omega_{4}(\bm{k})-\omega_{j}(\bm{k})]}
\bigg\}\;,\\
&&\frac{1}{N}\sum_{\bm{k}\sigma}\big[
\frac{n_{\mathrm{F}}(\epsilon_{+\bm{k}\sigma})}{\epsilon_{+\bm{k}\sigma}-\epsilon_{-\bm{k}\sigma}}(\epsilon_{+\bm{k}\sigma}-Q_{\bar{s}\bar{s}}^{f}\epsilon_{\bm{k}\bar{s}\sigma})
-\frac{n_{\mathrm{F}}(\epsilon_{-\bm{k}\sigma})}{\epsilon_{+\bm{k}\sigma}-\epsilon_{-\bm{k}\sigma}}(\epsilon_{-\bm{k}\sigma}-Q_{\bar{s}\bar{s}}^{f}\epsilon_{\bm{k}\bar{s}\sigma})
\big]\nonumber\\
&=&1 - \frac{1}{N}\sum_{\bm{k}}\bigg\{ \frac{\omega_{1}(\bm{k})\big[1+2n_{\mathrm{B}}[\omega_{1}(\bm{k})]\big]
\big[\omega_{1}(\bm{k})^2-U(\rho_{\bar{s}}+Q_{\bar{s}\bar{s}}^{X}\epsilon_{\bm{k}})\big] }
{\prod_{j\neq1}[\omega_{1}(\bm{k})-\omega_{j}(\bm{k})]}
+ \frac{\omega_{2}(\bm{k})\big[1+2n_{\mathrm{B}}[\omega_{2}(\bm{k})]\big]
\big[\omega_{2}(\bm{k})^2-U(\rho_{\bar{s}}+Q_{\bar{s}\bar{s}}^{X}\epsilon_{\bm{k}})\big] }
{\prod_{j\neq2}[\omega_{2}(\bm{k})-\omega_{j}(\bm{k})]} \nonumber\\
&+&\frac{\omega_{3}(\bm{k})\big[1+2n_{\mathrm{B}}[\omega_{3}(\bm{k})]\big]
\big[\omega_{3}(\bm{k})^2-U(\rho_{\bar{s}}+Q_{\bar{s}\bar{s}}^{X}\epsilon_{\bm{k}})\big] }
{\prod_{j\neq3}[\omega_{3}(\bm{k})-\omega_{j}(\bm{k})]}
+\frac{\omega_{4}(\bm{k})\big[1+2n_{\mathrm{B}}[\omega_{4}(\bm{k})]\big]
\big[\omega_{4}(\bm{k})^2-U(\rho_{\bar{s}}+Q_{\bar{s}\bar{s}}^{X}\epsilon_{\bm{k}})\big] }
{\prod_{j\neq4}[\omega_{4}(\bm{k})-\omega_{j}(\bm{k})]} \bigg\}\;. \label{Self-Eq-1-h}
\end{eqnarray}
\end{subequations}
At half filling, the left hand side of Eq.\eqref{Self-Eq-1-h} equals to $1$, while the first and second, the third and fourth momentum-dependent term at the right hand side offset each other, respectively, because of the bosonic energy dispersions $\omega_{j}(\bm{k})$ in Eq.\eqref{GF-Psi-X-2}, thus we have proved analytically that the equation \eqref{Self-Eq-1-h} is fulfilled automatically at half filling
with the vanishing chemical potential $\mu$ and Lagrange multiplier $h_{s}$, i.e., $\mu=h_{s}=0$. The parameters in Eq. \eqref{Self-Eq-1} are determined by the self-consistent calculation simultaneously
without any adjustable parameters, and the inverse of temperature $\beta$ and the lattice size $N$ in our calculation are set as 1000 and 400$\times$400, respectively.

\section{The derivation of the free energy of the system}\label{Free-En}

By Fourier transforming the Lagrange \eqref{Lagrange-LargeM} into momentum space the Lagrange is reexpressed as
\begin{eqnarray}\label{Lagrange-M2}
L &=&  \sum_{s=A,B}\sum_{\bm{k}}\big[\frac{|\partial_{\tau}X_{\bm{k}s}|^2-{h_{s}}^2}{U}+h_{s}+\rho_{s}(|X_{\bm{k}s}|^2-1)
+ \frac{h_{s}}{U}(\partial_{\tau}X_{\bm{k}s}X^\ast_{\bm{k}s}-\partial_{\tau}X^\ast_{\bm{k}s} X_{\bm{k}s})
+ f_{\bm{k}s}^{\dagger}(\partial_{\tau}-h_{s})f_{\bm{k}s} \big]\nonumber\\
&+& 3Nt[Q_{AB}^{X}Q_{BA}^{f}+\mathrm{H.C.}] - t\sum_{\bm{k}}\big[Q_{AB}^{X}X_{\bm{k}A}X_{\bm{k}B}^{\ast}V_{\bm{k}}^{\ast}
+ f_{\bm{k}A}^{\dagger}f_{\bm{k}B}Q_{BA}^{f}V_{\bm{k}} +\mathrm{H.C.}\big]\nonumber\\
&-& 6N\lambda\sum_{s=A,B}Q_{ss}^{X}Q_{ss}^{f} + \sum_{s=A,B}\sum_{\bm{k}}\big[ Q_{ss}^{f}\sum_{\sigma}\epsilon_{\bm{k}s\sigma}
f_{\bm{k}s\sigma}^{\dagger}f_{\bm{k}s\sigma} + Q_{ss}^{X}\epsilon_{\bm{k}}X_{\bm{k}s}X^{\ast}_{\bm{k}s} \big]\;,
\end{eqnarray}
then the action $S$ can be calculated directly from the above Lagrange, i.e.,
\begin{eqnarray}\label{ACTION-FULL-SADDLE}
S&=&\int_{0}^{\beta}L d\tau=\beta N\big\{\sum_{s}[\frac{-{h_{s}}^2}{U}+h_{s}-\rho_{s}]+3t[Q_{AB}^{X}Q_{BA}^{f}+\mathrm{H.C.}]-6\lambda\sum_{s=A,B}Q_{ss}^{X}Q_{ss}^{f} \big\}\nonumber\\
&+& \sum_{s=A,B}\sum_{\bm{k}i\nu}\big[\frac{\nu^2|X_{\bm{k}s}(i\nu)|^2}{U}+\rho_{s}|X_{\bm{k}s}(i\nu)|^2
- \frac{2h_{s}i\nu}{U}|X_{\bm{k}s}(i\nu)|^2\big]
+ \sum_{s=A,B}\sum_{\bm{k}i\omega\sigma}f_{\bm{k}s\sigma}^{\dagger}(i\omega)(-i\omega-h_{s})f_{\bm{k}s\sigma}(i\omega)\nonumber\\
&-& t\big[Q_{AB}^{X}\sum_{\bm{k}i\nu}X_{\bm{k}A}(i\nu)X_{\bm{k}B}^{\ast}(i\nu)V_{\bm{k}}^{\ast}
+\sum_{\bm{k}i\omega\sigma}f_{\bm{k}A\sigma}^{\dagger}(i\omega)f_{\bm{k}B\sigma}(i\omega)Q_{BA}^{f}V_{\bm{k}} +\mathrm{H.C.}\big]\nonumber\\
&+& \sum_{s=A,B}\big[ Q_{ss}^{f}\sum_{\bm{k}i\omega\sigma}\epsilon_{\bm{k}s\sigma}
f_{\bm{k}s\sigma}^{\dagger}(i\omega)f_{\bm{k}s\sigma}(i\omega) + Q_{ss}^{X}\sum_{\bm{k}i\nu}\epsilon_{\bm{k}}X_{\bm{k}s}(i\nu)X^{\ast}_{\bm{k}s}(i\nu) \big]\;,
\end{eqnarray}
which gives rise to the partition function $Z$ by integrating all quantum fields: $X$ and $f$
\begin{eqnarray}
Z&=& \int \prod\limits_{\bm{k}s=A,Bi\nu i\omega\sigma}dX_{\bm{k}s}(i\nu)dX^{\ast}_{\bm{k}s}(i\nu)df_{\bm{k}s\sigma}^{\dagger}(i\omega)df_{\bm{k}s\sigma}(i\omega)e^{-S}
= CZ_{X}Z_{f}\;,
\end{eqnarray}
where
\begin{subequations}
\begin{eqnarray}
Z_{f}&=&\prod\limits_{\bm{k}i\omega\sigma}\big\{-[-i\omega-h_{A}+Q_{AA}^{f}\epsilon_{\bm{k}A\sigma}][-i\omega-h_{B}+Q_{BB}^{f}\epsilon_{\bm{k}B\sigma}]
+t^2|Q_{BA}^{f}V_{\bm{k}}|^2 \big\} \;,\\
Z_{X}&=& \prod\limits_{\bm{k}i\nu}\frac{1}{|G_{X}(\bm{k},i\nu)^{-1}|} \;.
\end{eqnarray}
\end{subequations}
Then the free energy density $f=F/N$ is calculated as
\begin{eqnarray}\label{FE-Density}
f&=&-\frac{1}{N\beta}\ln Z = \sum_{s}[\frac{-{h_{s}}^2}{U}+h_{s}-\rho_{s}]+3t[Q_{AB}^{X}Q_{BA}^{f}+\mathrm{H.C.}]-6\lambda\sum_{s=A,B}Q_{ss}^{X}Q_{ss}^{f} \nonumber\\
&+& \frac{1}{N\beta}\sum_{\bm{k}i\nu}\ln\big[(\frac{\nu^2}{U}+\rho_{A} - \frac{2h_{A}i\nu}{U} + Q_{AA}^{X}\epsilon_{\bm{k}})(\frac{\nu^2}{U}+\rho_{B} - \frac{2h_{B}i\nu}{U} + Q_{BB}^{X}\epsilon_{\bm{k}})
- t^{2}|Q_{BA}^{X}V_{\bm{k}}|^{2} \big] \nonumber\\
&-& \frac{1}{N\beta}\sum\limits_{\bm{k}i\omega\sigma}\ln\big[-(-i\omega-h_{A}+Q_{AA}^{f}\epsilon_{\bm{k}A\sigma})(-i\omega-h_{B}+Q_{BB}^{f}\epsilon_{\bm{k}B\sigma})
+t^2|Q_{BA}^{f}V_{\bm{k}}|^2 \big] \nonumber\\
&=&  \sum_{s}[\frac{-{h_{s}}^2}{U}+h_{s}-\rho_{s}]+3t[Q_{AB}^{X}Q_{BA}^{f}+\mathrm{H.C.}]-6\lambda\sum_{s=A,B}Q_{ss}^{X}Q_{ss}^{f} \nonumber\\
&+& \frac{1}{N\beta}\sum_{\bm{k}j=1}^{4}\ln(1-e^{-\beta|\omega_{j\bm{k}}|})
- \frac{1}{N\beta}\sum\limits_{\bm{k}\sigma}\big[\ln(1+e^{-\beta\epsilon_{+\bm{k}\sigma}}) +\ln(1+e^{-\beta\epsilon_{-\bm{k}\sigma}}) \big]\;.
\end{eqnarray}

\section{The derivation of the Lagrange for a nanoribbon periodic in the zigzag direction and its Green's function}\label{Zigzag-Nanoribbon}

\begin{figure}[h!]
\centering
\includegraphics[scale=0.5]{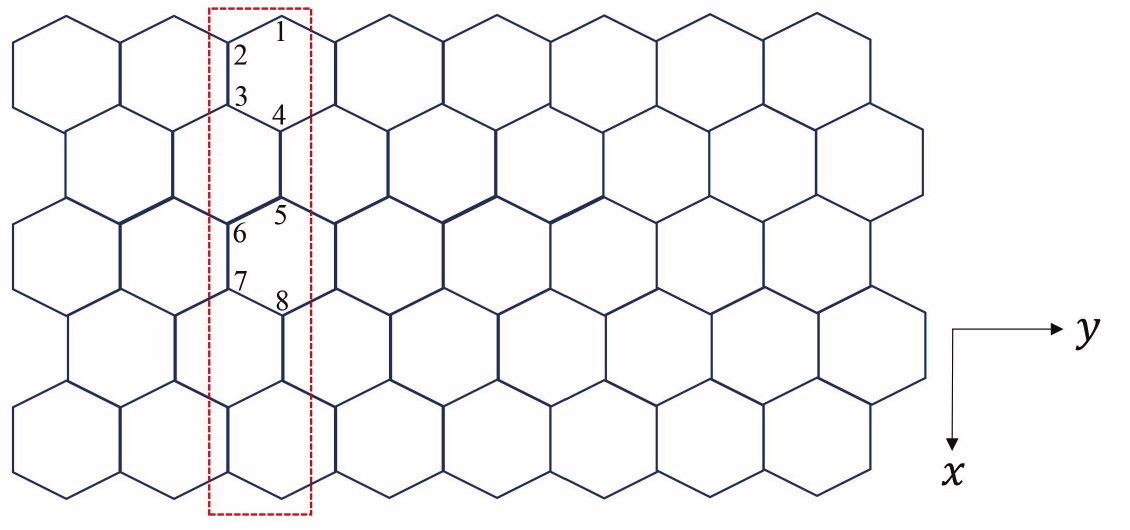}
\caption{(Color online) The illustration of a cylinder with an open boundary condition in the armchair ($x$) direction, while periodic in the zigzag ($y$) direction.
The unit cell in the zigzag direction is marked by a dashed red square where the numbers are used to label the atom locations.\label{Cylinder}}
\end{figure}
In this section we first extend the Lagrange (\ref{Lagrange-LargeM}) of the bulk system with periodic boundary condition to a cylindrical nanoribbon periodic in the zigzag($y$) direction
while with an open boundary condition in the armchair($x$) direction, which gives rise to the Lagrange of this cylinder
\begin{eqnarray}\label{Lagrange-LargeM-cyld1}
L&=&\sum_{i_xj_y\alpha}\big[\frac{|\partial_{\tau}X_{i_xj_y}^{\alpha}|^2-{h_{i_xj_y}}^2}{2U}+h_{i_xj_y}
+ \frac{h_{i_xj_y}}{2U}(\partial_{\tau}X_{i_xj_y}^{\alpha}{X_{i_xj_y}^{\alpha}}^\ast-\partial_{\tau}{X_{i_xj_y}^{\alpha}}^\ast X_{i_xj_y}^{\alpha})\big]+ \sum_{i_xj_y}\rho_{i_xj_y}(\sum_{\alpha}|X_{i_xj_y}^{\alpha}|^2-M)\nonumber\\
&+& \sum_{i_xj_y}f_{i_xj_y}^{\dagger}(\partial_{\tau}-h_{i_xj_y})f_{i_xj_y}+t\sum_{\langle i_xj_yi'_xj'_y\rangle}Q_{AB}^{X}(i_xj_y,i'_xj'_y)Q_{BA}^{f}(i'_xj'_y,i_xj_y)+\mathrm{H.C.} \nonumber\\
&-& t\sum_{\langle i_xj_yi'_xj'_y\rangle}\big[Q_{AB}^{X}(i_xj_y,i'_xj'_y)\sum_{\alpha}X_{i_xj_y}^{\alpha}{X_{i'_xj'_y}^{\alpha}}^{\ast}
+f_{i_xj_y}^{\dagger}f_{i'_xj'_y}Q_{BA}^{f}(i'_xj'_y,i_xj_y) +\mathrm{H.C.}\big]\nonumber\\
&-& \lambda\sum_{\langle\langle i_xj_yi'_xj'_y\rangle\rangle}Q_{ss}^{X}(i_xj_y,i'_xj'_y)Q_{ss}^{f}(i'_xj'_y,i_xj_y) \nonumber\\
&+& \lambda\sum_{\langle\langle i_xj_yi'_xj'_y\rangle\rangle}\big[ Q_{ss}^{f}(i'_xj'_y,i_xj_y)f_{i_xj_y}^{\dagger}
e^{i\tfrac{\pi}{2}\nu_{i_xj_yi'_xj'_y}\sigma^{z}}f_{i'_xj'_y} + Q_{ss}^{X}(i_xj_y,i'_xj'_y)\sum_{\alpha}X_{i_xj_y}^{\alpha}{X_{i'_xj'_y}^{\alpha}}^{\ast} \big]\;.
\end{eqnarray}
According to the previous work~\cite{Wagner24} where the site or bond dependent mean field parameters have been used to study the edge-state energy dispersion of a cylinder periodic in the zigzag direction
for the interacting Kane-Mele model in the chiral TBI state and topological Mott insulator state.
Their results shown in Fig. S4 indicate that though the parameters around two edges in the armchair direction are quantitatively different from those deep within the bulk,
the qualitative properties of the edge modes would not be altered
if the bulk parameters are used to study the physics of these edge states. Because the atoms labelled by odd numbers within a unit cell belong to sublattice A, while the others labelled by even numbers
belong to sublattice B, the above Lagrange can be rearranged as
\begin{eqnarray}\label{Lagrange-LargeM-cyld2}
L&=&\sum_{i_x\in {\rm odd} j_y\alpha}\big[\frac{|\partial_{\tau}X_{i_xj_y}^{\alpha}|^2-h_{A}^2}{2U}+h_{A}
+ \frac{h_{A}}{2U}(\partial_{\tau}X_{i_xj_y}^{\alpha}{X_{i_xj_y}^{\alpha}}^\ast-\partial_{\tau}{X_{i_xj_y}^{\alpha}}^\ast X_{i_xj_y}^{\alpha})\big]+
\sum_{i_x\in {\rm odd}j_y}\rho_{A}(\sum_{\alpha}|X_{i_xj_y}^{\alpha}|^2-M)\nonumber\\
&+&\sum_{i_x\in {\rm even} j_y\alpha}\big[\frac{|\partial_{\tau}X_{i_xj_y}^{\alpha}|^2-h_{B}^2}{2U}+h_{B}
+ \frac{h_{B}}{2U}(\partial_{\tau}X_{i_xj_y}^{\alpha}{X_{i_xj_y}^{\alpha}}^\ast-\partial_{\tau}{X_{i_xj_y}^{\alpha}}^\ast X_{i_xj_y}^{\alpha})\big]+
\sum_{i_x\in {\rm even}j_y}\rho_{B}(\sum_{\alpha}|X_{i_xj_y}^{\alpha}|^2-M)\nonumber\\
&+& \sum_{i_x\in {\rm odd}j_y}f_{i_xj_y}^{\dagger}(\partial_{\tau}-h_{A})f_{i_xj_y} + \sum_{i_x\in {\rm even}j_y}f_{i_xj_y}^{\dagger}(\partial_{\tau}-h_{B})f_{i_xj_y}\nonumber\\
&+& t\sum_{\langle i_xi'_xj_yj'_y\rangle}Q_{AB}^{X}Q_{BA}^{f}+\mathrm{H.C.}
- t\sum_{\langle i_xi'_xj_yj'_y\rangle}\big[Q_{AB}^{X}\sum_{\alpha}X_{i_xj_y}^{\alpha}{X_{i'_xj'_y}^{\alpha}}^{\ast}
+f_{i_xj_y}^{\dagger}f_{i'_xj'_y}Q_{BA}^{f} +\mathrm{H.C.}\big]\nonumber\\
&-& \lambda\sum_{\langle\langle i_xi'_x\in {\rm odd}j_yj'_y\rangle\rangle}Q_{AA}^{X}Q_{AA}^{f}
+ \lambda\sum_{\langle\langle i_xi'_x\in {\rm odd}j_yj'_y\rangle\rangle}\big[ Q_{AA}^{f}f_{i_xj_y}^{\dagger}
e^{i\tfrac{\pi}{2}\nu_{i_xj_yi'_xj'_y}\sigma^{z}}f_{i'_xj'_y} + Q_{AA}^{X}\sum_{\alpha}X_{i_xj_y}^{\alpha}{X_{i'_xj'_y}^{\alpha}}^{\ast} \big]\nonumber\\
&-& \lambda\sum_{\langle\langle i_xi'_x\in {\rm even}j_yj'_y\rangle\rangle}Q_{BB}^{X}Q_{BB}^{f}
+ \lambda\sum_{\langle\langle i_xi'_x\in {\rm even}j_yj'_y\rangle\rangle}\big[ Q_{BB}^{f}f_{i_xj_y}^{\dagger}
e^{i\tfrac{\pi}{2}\nu_{i_xj_yi'_xj'_y}\sigma^{z}}f_{i'_xj'_y} + Q_{BB}^{X}\sum_{\alpha}X_{i_xj_y}^{\alpha}{X_{i'_xj'_y}^{\alpha}}^{\ast} \big]\;.
\end{eqnarray}
Then the Lagrange \eqref{Lagrange-LargeM-cyld2} is expanded along the positive $x$ direction and the Fourier transformation in the $y$ direction is undertaken,
which directly gives rise to the spinon and chargon Green's function $G_{f\sigma}(i_x,i'_x,k_y,i\omega)$ and $G_{X}(i'_x,i_x,q_y,i\nu)$ defined on this cylinder
in the large $M$ limit. The electron Green's function defined on this cylinder can be expressed as
\begin{eqnarray}
G_{\sigma}(i_x,i'_x,k_y,i\omega)&=&\sum_{j_y-j'_y}\int_{0}^{\beta}d\tau e^{-ik_y(j_y-j'_y)+i\omega\tau}G_{f\sigma}(i_x,i'_x,j_y-j'_y,\tau)G_{X}(i'_x,i_x,j'_y-i_y,-\tau) \nonumber\\
&=& Z_{ss'}G_{f\sigma}(i_x,i'_x,k_y,i\omega)+\frac{1}{L\beta}\sum_{q_yi\nu}G_{f\sigma}(i_x,i'_x,k_y+q_y,i\omega+i\nu)G_{X}(i'_x,i_x,q_y,i\nu)\;,
\end{eqnarray}
where $s,s'=$ A or B correspond to $i_x,i'_x\in$ odd or even, respectively. Then the electron spectrum function defined on this cylinder can be obtained via the retarded electron Green's function, i.e.,
$A_{\sigma}(i_x,i'_x,k_y,\omega)=-2{\rm Im}G_{\sigma}(i_x,i'_x,k_y,\omega+i0^{+})$ where the infinitesimal analytical continuation width $0^{+}$ is set as $0.01$ in this work. For each $k_{y}$
and $\omega$, the overall electron spectral weight as a function of $k_y$ and $\omega$ is defined as
\begin{equation}
A_{\sigma}(k_y,\omega)=\frac{1}{L}\sum_{i_{x}}A_{\sigma}(i_x,i_x,k_y,\omega)\;,
\end{equation}
which indicates that the contributions to the above electron spectral weight come from both the bulk and two edges.

\begin{figure}[h!]
\centering
\includegraphics[scale=0.4]{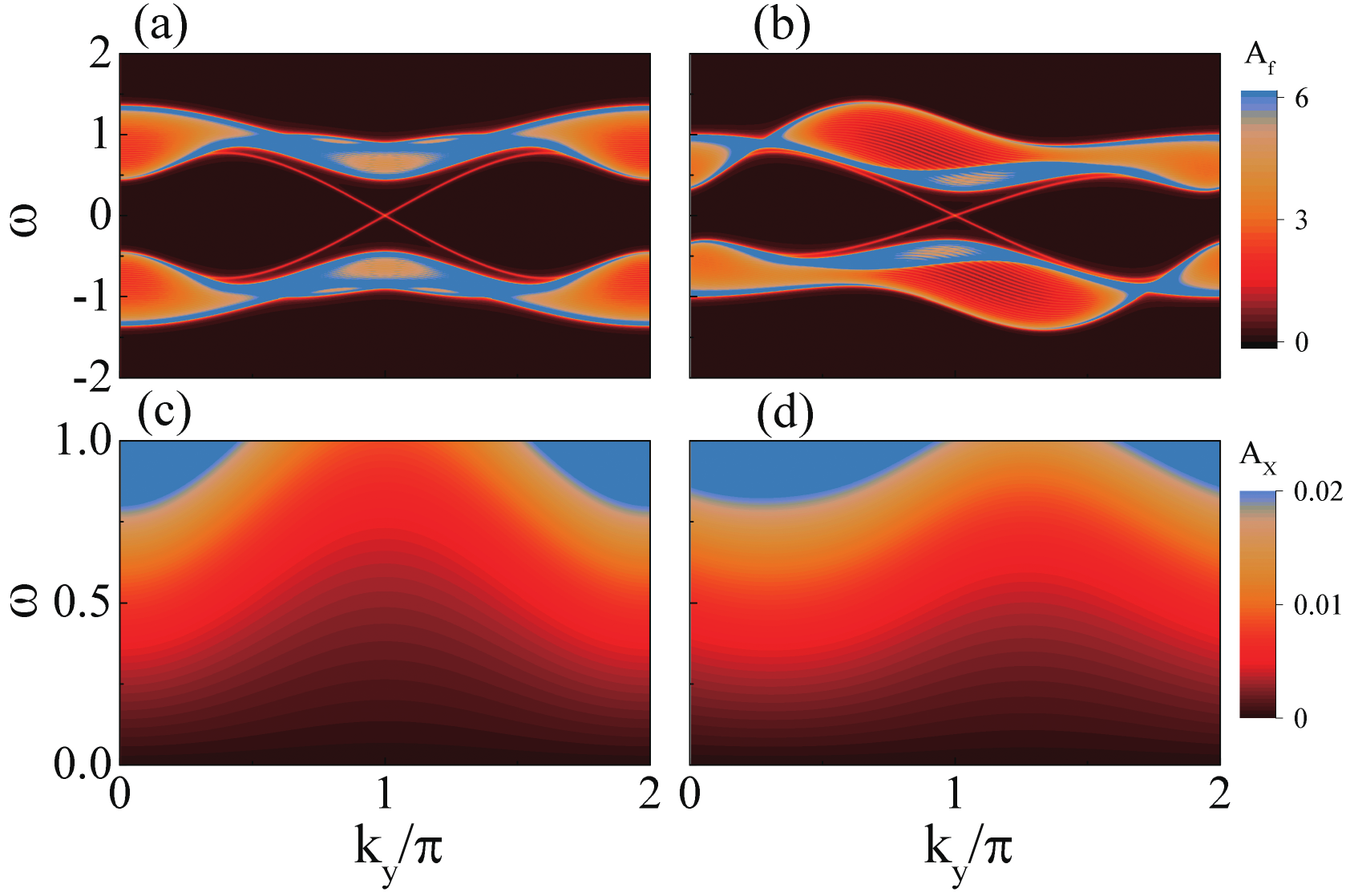}
\caption{(Color online) First row: the fermionic spinon spectrum function $A_{f\uparrow}(k_y,\omega)$ of the nanoribbon periodic along the zigzag direction at $\lambda=0.5$ and $U=2.7$
in the (a)chiral and (b)non-chiral TBI state; Second row: the corresponding bosonic chargon spectrum function $A_{X}(k_y,\omega)$ in the (c)chiral and (d)non-chiral TBI state.\label{EDGE-U27-PSI-X}}
\end{figure}
In the main text, we have studied the electron spectrum function of the nanoribbon periodic in the zigzag direction as shown in Figs.\ref{Edge-State} and \ref{EDGE-U27SPINUP-DN}, where the results show that the spontaneously broken chiral symmetry driven by the intrinsic SOC in the presence of electron coulomb interaction leads to the asymmetric bulk spectral weight distribution and edge-mode energy dispersion with respect to the time reversal invariant point $k_y=\pi$.
In particular, the inequivalent edge states at opposite boundaries in the zigzag direction give rise to a net spin current across the system which benefits the design of advanced spintronics devices.
However, we find that the crossing point at $k_y=\pi$ of two edge modes for each spin deviates from zero energy, which thus exhibits significant difference from the noninteracting system.
Thus to disclose the underlying mechanism for this deviation, we further investigate the spinon and chargon spectrum function of the nanoribbon periodic along the zigzag direction at $\lambda=0.5$ and $U=2.7$
for the chiral and non-chiral TBI state in the first and second column of Fig.\ref{EDGE-U27-PSI-X}, respectively.
As shown in Fig.\ref{EDGE-U27-PSI-X}(a) and (b), the spectrum function of the fermionic spinon in the chiral and non-chiral TBI state exhibits notable edge-state energy dispersion.
Most importantly, the spontaneously broken chiral symmetry in the non-chiral TBI state leads to the asymmetrical
energy dispersion with respect to the time reversal invariant point $k_y=\pi$, which then gives rise to the asymmetrical electron spectral weight distribution shown in Fig.\ref{Edge-State}
as a result of charge-spin recombination. As shown in Fig.\ref{EDGE-U27-PSI-X}(c) and (d),
compared to the gapless bosonic chargon excitations in $X$ for the bulk system periodic in $\bm{a}_{1}$ and $\bm{a}_{2}$ direction, the open boundary condition in the armchair direction gives rise to
a finite energy gap demonstrated by the vanishing spectral weight around $k_{y}=0$ and $2\pi$, which then leads to that the crossing point of the electron edge-mode dispersion deviates from zero energy.
In addition, the spontaneously broken chiral symmetry gives rise to the asymmetric spectral weight distribution for the bosonic field $X$ with respect to $k_y=\pi$ in the non-chiral TBI state.

\end{widetext}

\bibliography{BIBINSOC}

\end{document}